**ECLIPSE-X: Plate-Scale Calibration as the Binding Constraint on Relativistic Astrometry with Spacecraft Navigation**

**Pedro J. Llanos**

Department of Applied Aviation Sciences, College of Aviation, Embry-Riddle Aeronautical University, 1 Aerospace Boulevard, Daytona Beach, FL 32114, United States

Corresponding author: llanosp@erau.edu

ORCID: 0000-0002-8393-089X

**Abstract**

Precision stellar astrometry near the Sun can constrain the parametrized post-Newtonian (PPN) parameter γ through gravitational light deflection. This study develops ECLIPSE-X, a localized multi-frame estimation architecture in which one persistent relativistic parameter is estimated simultaneously with frame-dependent spacecraft line-of-sight and roll states, exploiting the differing spatial signatures of deflection, field translation and roll together with the temporal distinction between a persistent γ and independently varying frame-local navigation. The objective is not a competitive determination of γ, which is independently constrained to substantially higher precision, but an assessment of whether such a joint relativistic-navigation solution and its formal covariance remain statistically valid under realistic astrometric model error. A deterministic synthetic field of 250 stars spanning 1.22–8 apparent solar radii is observed over 40 frames at 5-s cadence; weighted least squares estimates γ together with three navigation states per frame, and Schur-complement marginalization gives the formal uncertainty in γ after navigation coupling. The nominal solution recovers $\hat{\gamma}$ = 1.0001198167 with $\sigma_\gamma$ = 3.075104 × $10^{-4}$, and a 1000-realization matched-model ensemble confirms statistical calibration. Introducing astrometric perturbations unknown to the estimator identifies plate-scale mismatch as the dominant failure mechanism: a 300-ppm scale error alone inflates the empirical-to-formal uncertainty ratio to 7126.6 with zero 1σ and 2σ coverage, against 2.18 for catalog-coordinate mismatch and 1.12 for a frame-coherent radial perturbation, while every realization converges numerically — numerical convergence does not guarantee statistical validity. Augmenting the state vector with a plate-scale parameter restores consistency, at a fixed 21.2 per cent cost in the precision of γ set by the correlation between the two persistent parameters at this observing geometry. A sweep in calibration uncertainty shows the unaugmented estimator remains covariance-consistent only below $\sigma_p$ = 2.7884 × $10^{-8}$, so the augmented state is a requirement rather than a refinement for this class of near-Sun relativistic astrometric estimator.

## 1. Introduction

Solar gravitational light deflection is one of the classical observational consequences of general relativity (GR). The eclipse observations of 1919 provided the first widely recognized measurement of the apparent displacement of background stellar positions produced by the solar gravitational field [1]. Within the parametrized post-Newtonian formalism, the parameter γ quantifies spatial curvature produced per unit rest mass and has the GR value [2-4]

$$\gamma_{GR} = 1 \tag{1}$$

For a light ray grazing the Sun, GR predicts a leading-order deflection of approximately 1.75 arcsec [2,3]. Subsequent radio-interferometric and spacecraft measurements improved experimental constraints substantially. The Cassini solar-conjunction experiment obtained $\gamma - 1 = (2.1 \pm 2.3) \times 10^{-5}$ [5], which remains among the tightest single-experiment constraints on $\gamma$.

Relativistic light propagation is intrinsic to precision space astrometry. Klioner [6] developed a practical relativistic astrometric framework incorporating gravitational light propagation, observer motion, aberration, parallax, and proper motion, while the International Astronomical Union relativistic reference-system framework provides the associated coordinate and metrological basis [7]. Gaia extended the estimation problem to simultaneous determination of stellar parameters, spacecraft attitude, instrument calibration, and global quantities [8,9]. Hobbs et al. [10] specifically investigated determination of PPN $\gamma$ while solving simultaneously for stellar parameters and satellite attitude. Joint estimation of relativistic and spacecraft-dependent quantities therefore has established precedent.

Dedicated near-Sun concepts have investigated complementary approaches. The Laser Astrometric Test of Relativity (LATOR) proposed precision solar-grazing optical interferometry [11,12], whereas the Gamma Astrometric Measurement Experiment (GAME) investigated differential stellar astrometry close to the solar limb [13,14].

ECLIPSE-X examines a different localized estimation architecture. A single relativistic parameter persists throughout a multi-frame observation sequence, while transverse LOS and roll states are independently estimated in each frame. The resulting problem contains two complementary sources of observability: spatial differentiation among radial gravitational deflection, translational LOS displacement, and tangential roll displacement; and temporal differentiation between one persistent relativistic state and frame-local navigation states.

From a spacecraft-navigation perspective, the significance of this formulation is not limited to estimating the PPN parameter $\gamma$. The same stellar image sequence is used to separate a persistent physical signal from time-varying spacecraft LOS and roll motion. This makes the problem directly relevant to precision optical navigation: a calibration error that projects onto the same measurement geometry as the desired physical or navigation states can bias both while the numerical estimator still appears to converge. Accordingly, ECLIPSE-X tests not only whether $\gamma$ can be recovered, but also whether the accompanying navigation solution and its formal covariance remain trustworthy when coherent astrometric calibration errors are present.

A second objective is to determine how this architecture behaves when the assumed astrometric geometry differs from physical truth. This distinction is important because formal covariance describes uncertainty conditional on the estimator model. An estimator can therefore converge numerically while remaining statistically invalid if coherent calibration errors are absent from its state or noise model.

The marginalized uncertainty attainable from a localized 250-star, 40-frame sequence is $\sigma_\gamma \approx 3 \times 10^{-4}$, more than an order of magnitude larger than the Cassini constraint, and no competitive determination of $\gamma$ is claimed here. That relationship is what makes $\gamma$ a suitable test parameter rather than a limitation of the study. Because its true value is independently established to far better precision than this architecture can reach, any dispersion or covariance inconsistency recovered from the Monte Carlo ensembles is attributable to the estimator and its calibration assumptions rather than to residual uncertainty in the underlying physics. The quantity under examination is the trustworthiness of the solution, not the value of $\gamma$.

The contributions of this study are correspondingly diagnostic rather than metrological: (1) formulation of a localized persistent-global/frame-local relativistic-navigation estimator; (2) explicit marginalization of 120 frame-dependent navigation nuisance states in determining information on $\gamma$; (3) Monte Carlo verification of formal covariance calibration under matched astrometric assumptions; (4) controlled truth-versus-model experiments under hidden astrometric mismatch; and (5) source-isolation ablation ranking catalog-coordinate, plate-scale, and frame-coherent radial errors, which identifies plate-scale mismatch as the dominant failure mechanism and traces that dominance to the radial structure of the error mode rather than to its amplitude. To our knowledge, the propagation of unmodeled instrument-calibration error into the recovered PPN parameter and its formal covariance has not previously been quantified for a joint relativistic-astrometric and navigation solution. The consequence for mission design is that plate-scale knowledge belongs in the science error budget and not only in the instrument calibration plan.

## 2. Mathematical and Computational Method

### 2.1 Solar Angular Geometry

Physical interpretation. This subsection converts the apparent position of a star on the sky into the impact geometry required by the relativistic light-deflection model. The observer sees the solar radius as an angle, while the gravitational model uses the closest-approach distance $b_i$ of the light ray to the Sun. Equations (2)-(6) provide that bridge. The small-angle approximation is appropriate because all angles in the local near-Sun tangent plane are much smaller than one radian. In the MATLAB implementation these quantities are formed from C.Rsun, C.AU, catalog.rho_rad, catalog.impact_m, catalog.eRx, and catalog.eRy inside the physical-constants block and makeSyntheticStarCatalog().

At observer-to-Sun distance $D_\odot$, the apparent angular solar radius is approximated by the standard small-angle relation

$$\rho_\odot \simeq \frac{R_\odot}{D_\odot} \tag{2}$$

For stellar source i, define its angular separation in apparent solar radii as

$$q_i = \frac{\rho_i}{\rho_\odot} \tag{3}$$

The exact physical impact parameter is

$$b_i = D_\odot \tan\rho_i \tag{4}$$

which, under the small-angle approximation, becomes [2,3,6]

$$b_i \simeq D_\odot \rho_i \simeq q_i R_\odot \tag{5}$$

The corresponding radial unit vector in the local tangent plane is

$$\hat{\mathbf{r}}_i = \frac{1}{\rho_i}\begin{bmatrix}\theta_{x,i}^0 \\ \theta_{y,i}^0\end{bmatrix} \tag{6}$$

Figure 1 summarizes the near-Sun observation geometry underlying Eqs. (2)-(6), connecting the physical Sun-observer configuration to the tangent-plane coordinates used by the relativistic navigation estimator.

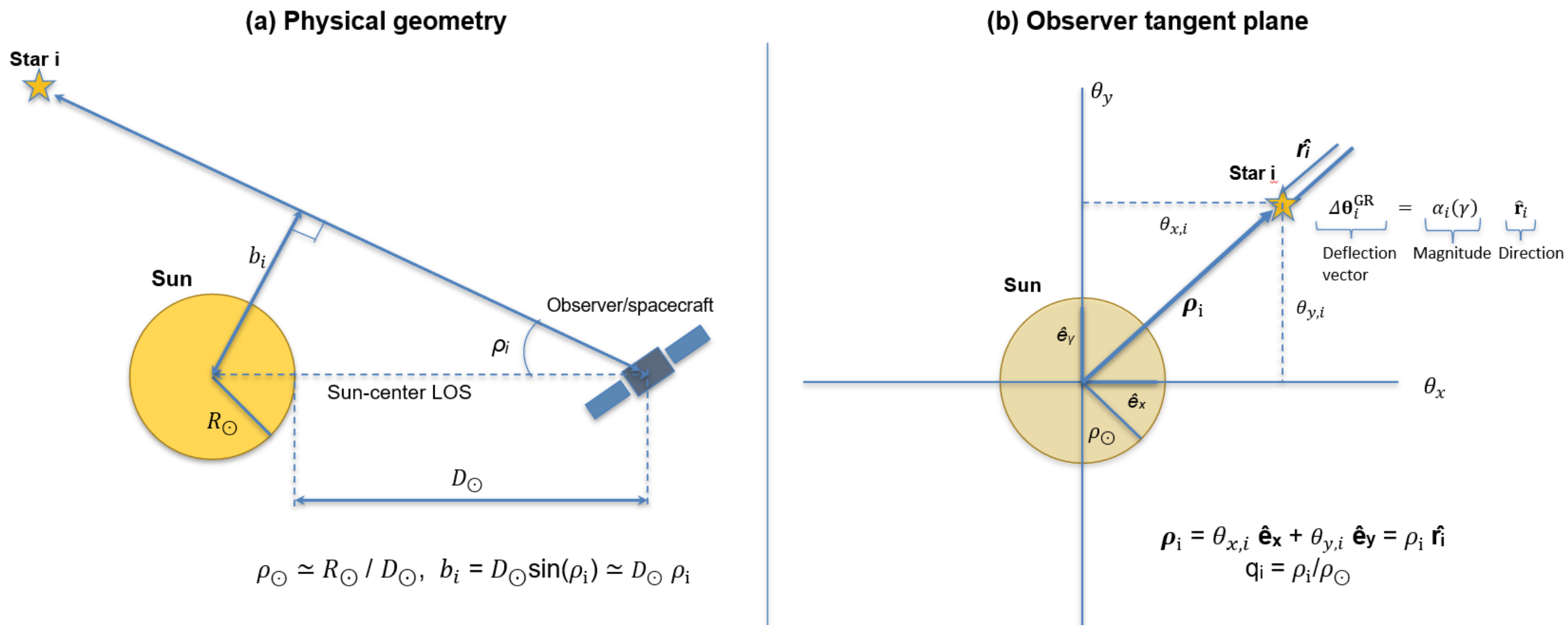

**Fig. 1. Near-Sun astrometric geometry used in ECLIPSE-X. (a) Physical observation geometry relating the solar radius R⊙, observer–Sun distance D⊙, stellar angular separation ρᵢ, and corresponding impact parameter bᵢ. (b) Observer tangent-plane representation in which the stellar angular-position vector is resolved into components $\theta_{x,i}^{0}$ and $\theta_{y,i}^{0}$ along the basis vectors ê_x and ê_y. The radial unit vector $\hat{\mathbf{r}}_i$ defines the direction of the relativistic deflection $\Delta\theta_i^{GR}$, while $q_i$ expresses stellar separation in apparent solar radii. Geometry is schematic and not to scale.**

### 2.2 Deterministic Synthetic Stellar Field

Why these variables are used. The variables $u_i$ and $v_i$ are dimensionless uniform random numbers used only to construct a reproducible test geometry; they are not measured stellar properties. The variable $q_i$ is the star-Sun angular separation expressed in apparent solar radii, and $\varphi_i$ specifies azimuth around the Sun. Using $q_i$ rather than angle in radians makes the near-limb geometry immediately interpretable: $q_i = 1$ corresponds approximately to the apparent solar limb. The range 1.22-8 $R_\odot$ deliberately includes both strongly deflected inner-field stars and more weakly deflected outer-field stars so that the estimator has radial leverage. In code this is cfg.minSolarSep_Rsun, cfg.maxSolarSep_Rsun, the exponent 1.5, and the arrays rho_Rsun, phi, thetaX, and thetaY in makeSyntheticStarCatalog().

The synthetic field is intended to provide controlled estimator geometry rather than reproduce an astrophysical stellar-density distribution. MATLAB's Mersenne-Twister pseudorandom generator is initialized using

$$\text{rng(14018,twister)} \tag{7}$$

where the seed has no physical significance and is retained solely for reproducibility. For independent random variables

$$u_i, v_i \sim \mathcal{U}(0,1) \tag{8}$$

stellar radial separation is generated according to

$$q_i = 1.22 + (8 - 1.22)u_i^{3/2} \tag{9}$$

Because $b_i \simeq D_\odot\rho_i$, $\rho_i = q_i\rho_\odot$, and $\rho_\odot \simeq R_\odot/D_\odot$, the stellar impact parameter can be written as $b_i \simeq q_i R_\odot$

Since the leading-order solar gravitational deflection is inversely proportional to impact parameter, it follows that

$$\alpha_i \propto q_i^{-1}, \quad \frac{\alpha(1.22)}{\alpha(8)} = \frac{8}{1.22} \simeq 6.56 \tag{10}$$

Thus, a star at $q_i = 1.22$, corresponding to 1.22 apparent solar radii, carries approximately 6.56 times the first-order gravitational-deflection amplitude of a star at $q_i = 8$. This radial contrast provides useful leverage for distinguishing the persistent relativistic signature from frame-local navigation errors.

Azimuth is sampled uniformly,

$$\phi_i = 2\pi v_i \tag{11}$$

and

$$\rho_i = q_i \rho_\odot \tag{12}$$

Nominal tangent-plane coordinates are therefore

$$\theta_{x,i}^0 = \rho_i \cos\phi_i, \qquad \theta_{y,i}^0 = \rho_i \sin\phi_i \tag{13}$$

where the superscript denotes the nominal catalog tangent-plane coordinate prior to the application of relativistic deflection, frame-dependent navigation perturbations, measurement noise, and astrometric model mismatch.

Equation (9) is an ECLIPSE-X numerical sampling relation rather than an astrophysical stellar-distribution law. The inner boundary $q_{\min} = 1.22$ retains strong relativistic sensitivity while avoiding placement directly on the apparent solar limb. The outer boundary $q_{\max} = 8$ extends the field into a region where gravitational displacement is substantially weaker, providing spatial leverage between relativistic and navigation signatures. The exponent 3/2 moderately weights the synthetic sample toward the high-sensitivity inner field because $u^{3/2} < u$ for $0 < u < 1$. It is a numerical design choice rather than an optimized or physically inferred exponent. The simulation contains $N_s = 250$ stars. For the fixed 250-star seeded realization used in this study, the resulting catalog spans $q_i = 1.221\text{-}7.919$.

### 2.3 PPN Solar Gravitational Deflection

The PPN parameter γ multiplies the part of the deflection associated with spatial curvature. General relativity predicts $\gamma = 1$. The factor $2GM_\odot/(c^2 b_i)$ sets the geometric strength of the solar gravitational deflection for star $i$; multiplication by $(1 + \gamma)$ gives the leading-order PPN deflection magnitude. The radial unit vector $\hat{r}_i$ then assigns that scalar deflection to the two tangent-plane coordinates. In MATLAB, q=2*C.G*C.Msun./(C.c^2*b), while catalog.gx and catalog.gy are the partial derivatives of the two measured coordinates with respect to $\gamma$. These derivatives are the sensitivity directions used by the estimator.

The leading-order PPN solar light-deflection magnitude is [2,3,6]

$$\alpha_i(\gamma) = (1+\gamma)\frac{2GM_\odot}{c^2 b_i} \tag{14}$$

For GR, $\gamma = 1$, and therefore

$$\alpha_i^{\mathrm{GR}} = \frac{4GM_\odot}{c^2 b_i} \tag{15}$$

At $b_i = R_\odot$, the constants used in the implementation give

$$\alpha_{\mathrm{limb}} = 1.751243 \text{ arcsec} \tag{16}$$

The two-dimensional tangent-plane deflection is

$$\Delta\boldsymbol{\theta}_i^{\mathrm{GR}} = \alpha_i(\gamma)\hat{\mathbf{r}}_i \tag{17}$$

Differentiation with respect to γ yields

$$\mathbf{g}_i = \frac{\partial\Delta\boldsymbol{\theta}_i^{\mathrm{GR}}}{\partial\gamma} = \frac{2GM_\odot}{c^2 b_i}\hat{\mathbf{r}}_i \tag{18}$$

where $\mathbf{g}_i$ is the 2-D astrometric sensitivity vector describing how the predicted tangent-plane position of star changes with the relativistic parameter, $\gamma$.

**2.4 Observation Sequence and Navigation Truth**

A single image can contain strong geometric coupling between a relativistic displacement and spacecraft pointing error. The 40-frame sequence forces gamma to remain common to every frame while LOS and roll are allowed to change with time. The sinusoidal histories are controlled stress-test trajectories, not predictions of spacecraft motion. Their amplitudes are cfg.losAmpX_rad=2.0e-7 rad, cfg.losAmpY_rad=1.5e-7 rad, and cfg.rollAmp_rad=2.0e-6 rad. The MATLAB truth arrays truth.dThetaX, truth.dThetaY, and truth.dPsi implement Eqs. (22)-(26).

Frames are sampled according to

$$t_k = (k-1)\Delta t \tag{19}$$

with Δt = 5 s and $N_f$ = 40.

$$\Delta t = 5\ \mathrm{s}, \qquad N_f = 40 \tag{20}$$

The resulting duration is

$$T = (N_f - 1)\Delta t = 195\ \mathrm{s} \tag{21}$$

The simulated transverse LOS histories are sinusoidal in both axes, providing a smooth, bounded, time-varying pointing disturbance against which the estimator must separate the persistent signal, $\gamma$.

$$\delta\theta_x^{\mathrm{true}}(t) = 2.0\times10^{-7}\sin\left(\frac{2\pi t}{T}\right)\ \mathrm{rad} \tag{22}$$

Similarly, the corresponding y-LOS sinusoid uses a different amplitude/phase history so the two transverse pointing components are not artificially identical or perfectly coupled:

$$\delta\theta_y^{\mathrm{true}}(t) = 1.5\times10^{-7}\cos\left(\frac{2\pi t}{T}+0.35\right)\ \mathrm{rad} \tag{23}$$

Roll is prescribed as a controlled time-dependent rotation about the LOS, providing a third independent frame-local navigation disturbance

$$\delta\psi^{\mathrm{true}}(t) = 2.0\times10^{-6}\sin\left(\frac{4\pi t}{T}+0.6\right)\ \mathrm{rad} \tag{24}$$

For a small roll angle, $\delta\psi_k$, the first-order coordinate perturbation follows the standard infinitesimal 2D rotation

$$\Delta\boldsymbol{\theta}_{i,k}^{\mathrm{roll}} = \delta\psi_k\mathbf{J}\boldsymbol{\theta}_i^0 \tag{25}$$

where, **J** is the 2D skew-symmetric rotation generator

$$\mathbf{J} = \begin{bmatrix} 0 & -1 \\ 1 & 0 \end{bmatrix} \tag{26}$$

The cadence, frame count, and prescribed navigation histories are controlled stress-test inputs rather than specifications for a particular spacecraft sensor.

### 2.5 Persistent-Global/Frame-Local State Architecture

State interpretation. The estimator intentionally separates one global scientific quantity from local navigation nuisance parameters. $\gamma$ is persistent because the same physical PPN parameter applies to the entire observation sequence. In contrast, $\delta\theta_{x,k}$, $\delta\theta_{y,k}$, and $\delta\psi_k$ are allowed to vary independently from frame to frame because spacecraft pointing need not remain constant. With 40 frames, this produces 120 navigation nuisance states but only one persistent relativistic state. The code stores the recovered frame states in est.dThetaXHat, est.dThetaYHat, and est.dPsiHat and the persistent solution in est.gammaHat.

The frame-local navigation state groups the three navigation errors that can change independently in each frame: two transverse LOS offsets and one roll offset:

$$\mathbf{n}_k = [\delta\theta_{x,k} \quad \delta\theta_{y,k} \quad \delta\psi_k]^T \tag{27}$$

The complete state combines the one persistent parameter $\gamma$ with all $N_f$ frame-local navigation vectors, because $\gamma$ is common to the experiment while spacecraft pointing can change from frame to frame:

$$\mathbf{x} = \begin{bmatrix}\gamma & \mathbf{n}_1^T & \mathbf{n}_2^T & \cdots & \mathbf{n}_{N_f}^T\end{bmatrix}^T \tag{28}$$

Equation (28) collects the complete estimation state. The first element, $\gamma$, is the single persistent scientific parameter shared by all frames. Each vector $n_k$ contributes three frame-local nuisance parameters, $\delta\theta_{x,k}$, $\delta\theta_{y,k}$, and $\delta\psi_k$, so the full state contains one global parameter plus $3N_f$ navigation parameters. The superscript $^T$ denotes transpose and is used only to assemble the individual column-state blocks into one column vector.

The relativistic state persists across all frames, where the relativistic parameter is physically invariant across the observation sequence rather than being re-estimated as an independent quantity in every frame.

$$\gamma_k = \gamma, \qquad k = 1, \ldots, N_f \tag{29}$$

For $N_f = 40$, we get $\gamma$ plus 120 frame-local navigation nuisance states:

$$N_x = 1 + 3N_f = 121 \tag{30}$$

### 2.6 Measurement Noise

Measurement interpretation and parameter rationale. The quantity sigma_cent=0.0200 pixel is the assumed one-standard-deviation centroid-location uncertainty in the detector image. It describes how precisely the center of a stellar point-spread function is in pixel coordinates. The estimator itself operates in angular coordinates, so the simulation also specifies sigma_theta=5.0e-8 rad as the corresponding one-standard-deviation angular LOS measurement uncertainty. Their ratio, 2.5e-6 rad pixel^-1 (approximately 0.516 arcsec pixel^-1), is an implied angular conversion used to relate image-plane precision to sky-angle precision; it is not presented as a measured calibration of a specific flight camera. These controlled values are stored as cfg.centroidSigma_pix and cfg.angularSigma_rad. The actual measurement covariance used by the estimator is set by obs.sigmaRad=cfg.angularSigma_rad*sigmaScale in simulateSequence().

The variable sigmaScale is deliberately separate from plateScaleSigma. sigmaScale multiplies random measurement noise and therefore changes the assumed standard deviation of each centroid-derived angular observation. plateScaleSigma instead produces a coherent geometric calibration error that stretches or contracts the truth star field while remaining hidden from the estimator. Keeping these quantities separate is essential: one represents random measurement precision, whereas the other represents structured model error.

The controlled baseline centroid uncertainty with which a stellar centroid is assumed to be located on the detector is a simulation precision assumption (not a derived physical constant):

$$\sigma_{\text{cent}} = 0.0200 \text{ pixel} \tag{31}$$

$\sigma_\theta$ is the corresponding controlled $1\sigma$ uncertainty used by the estimator for each tangent-plane coordinate.

$$\sigma_\theta = 5.0 \times 10^{-8} \text{ rad} \tag{32}$$

The implied angular scale is the ratio between the assumed pixel centroid precision to the assumed angular measurement precision:

$$s_{\text{pix}} = \frac{\sigma_\theta}{\sigma_{\text{cent}}} = 2.5 \times 10^{-6} \text{ rad pixel}^{-1} \tag{33}$$

Physical camera interpretation of plate scale. Equation (33) expresses the angular scale implied by the adopted centroid and angular uncertainties, but plate scale also has a direct optical meaning. For an idealized focal-plane camera, a source at field angle θ is mapped to detector coordinate x through the focal length f.

$$x = f \tan\theta \tag{33a}$$

For sufficiently small field angles, tan θ ≈ θ, so the focal-plane mapping becomes approximately linear:

$$x \simeq f\theta, \quad \theta \simeq \frac{x}{f} \tag{33b}$$

If p denotes the physical detector pixel pitch, one pixel corresponds approximately to the angular increment

$$s_{pix} \simeq \frac{p}{f} \quad [rad\ pixel^{-1}] \tag{33c}$$

Thus plate scale is fundamentally an instrument-calibration quantity: it depends on the mapping between physical detector distance and sky angle. Changes in effective focal length, detector geometry, thermoelastic alignment, focus, or optical distortion can therefore alter the true angular scale even when stellar centroids are measured with very small random error. This distinction is essential in ECLIPSE-X: $\sigma_\theta$ describes random angular measurement noise, whereas $s_p$ describes a coherent realization-wide error in the angular mapping itself.

Measurement noise is modeled as

$$\boldsymbol{\epsilon}_{i,k} \sim \mathcal{N}(\mathbf{0}, \sigma_\theta^2 \mathbf{I}_2) \tag{34}$$

So that each star has zero-mean Gaussian angular error with equal variance $\sigma_\theta^2$ in x and y directions, and the identity covariance assumes these two coordinate errors are uncorrelated.

In Eq. (34), $\varepsilon_{i,k}$ is the two-component measurement-noise vector associated with star $i$ in frame $k$; **0** is the two-component zero mean; $\sigma_\theta$ is the one-standard-deviation angular error applied to each tangent-plane coordinate; and $\mathbf{I}_2$ is the 2×2 identity matrix. The covariance $\sigma_\theta^2\mathbf{I}_2$ therefore assumes equal x- and y-coordinate variance and no correlation between the two components in the baseline model.

### 2.7 Nominal Measurement Model

Reading Eq. (35). Each measured star coordinate is the sum of four conceptually different contributions: its nominal catalog position, solar gravitational deflection, common transverse LOS translation, and the small displacement caused by spacecraft roll, plus random measurement noise. This decomposition is important because the estimator succeeds only to the extent that these signatures are distinguishable across the stellar field and across time. The same construction appears in simulateSequence(), where xGR/yGR form the relativistically deflected truth position and xNav/yNav add LOS and roll.

Under matched truth and estimator geometry,

$$\mathbf{z}_{i,k} = \boldsymbol{\theta}_i^0 + \Delta\boldsymbol{\theta}_i^{\mathrm{GR}}(\gamma) + \begin{bmatrix}\delta\theta_{x,k}\\ \delta\theta_{y,k}\end{bmatrix} + \delta\psi_k \mathbf{J}\boldsymbol{\theta}_i^0 + \boldsymbol{\epsilon}_{i,k} \tag{35}$$

where $\boldsymbol{\theta}_i^0$ is the nominal catalog position of star i in the local tangent plane; $\Delta\theta_i^{\mathrm{GR}}(\gamma)$ is its solar gravitational-deflection vector for the current value of γ; $[\delta\theta_{x,k}, \delta\theta_{y,k}]^{\mathrm{T}}$ is the common transverse LOS translation for frame k; $\delta\psi_k \mathbf{J}\boldsymbol{\theta}_i^0$ is the first-order tangential displacement caused by frame roll so it rotates each stellar position tangentially about the field origin; and $\boldsymbol{\epsilon}_{i,k}$ is the random centroid-derived angular measurement error. The indices i and k distinguish the star and frame, respectively. Only γ is persistent across the complete sequence; the LOS and roll terms are re-estimated independently in every frame. The spatial modes differ fundamentally: gravitational deflection is radial and approximately $b^{-1}$-dependent, LOS error is translational, and roll displacement is tangential and position dependent.

### 2.8 Truth-versus-Model Astrometric Mismatch

Three structured mismatch classes are used throughout the remainder of the paper. CAT denotes catalog-coordinate mismatch: each star is assigned a small realization-wide coordinate offset that is unknown to the estimator. SCALE denotes plate-scale mismatch: the complete truth field is multiplied by one realization-wide fractional scale factor, so displacement grows in proportion to distance from the field center. RAD denotes frame-coherent radial mismatch: during a given frame all stars receive an additional displacement of common scalar magnitude along their own local Sun-to-star radial directions. CAT therefore changes individual star locations irregularly; SCALE expands or contracts the field; RAD produces a coherent radial shift of approximately equal angular magnitude within a frame.

These are deliberately truth-versus-model tests. In simulateSequence(), CAT is generated by dCatX and dCatY, SCALE by plateScaleError and plateScaleTruth, and RAD by radialSys and obs.radialSysHistory. The estimator is intentionally not given these quantities: obs.thetaXModel and obs.thetaYModel remain equal to the nominal catalog coordinates. Consequently, the experiments test model adequacy rather than merely increasing the covariance of a correctly specified estimator.

Catalog-coordinate errors are assumed to be zero-mean Gaussian perturbations, a controlled model for star-specific catalog-coordinate uncertainty $\sigma_{cat}$:

$$\delta\theta_{x,i}^{\mathrm{cat}}, \delta\theta_{y,i}^{\mathrm{cat}} \sim \mathcal{N}(0, \sigma_{\mathrm{cat}}^2) \tag{36}$$

The perturbed stellar coordinate is

$$\widetilde{\boldsymbol{\theta}}_i = \boldsymbol{\theta}_i^0 + \begin{bmatrix}\delta\theta_{x,i}^{\mathrm{cat}}\\ \delta\theta_{y,i}^{\mathrm{cat}}\end{bmatrix} \tag{37}$$

In Eq. (37), the tilde identifies the catalog position used to generate the perturbed truth geometry before plate scale is applied. The two additive components $\delta\theta_{x,i}^{\mathrm{cat}}$ and $\delta\theta_{y,i}^{\mathrm{cat}}$ are independent star-specific coordinate errors drawn once for each Monte Carlo realization and then retained for all frames of that realization.

A realization-wide fractional plate-scale perturbation is generated as

$$s_p \sim \mathcal{N}\left(0, \sigma_p^2\right) \tag{38}$$

giving physical truth coordinates

$$\boldsymbol{\theta}_i^{\text{true}} = (1 + s_p)\tilde{\boldsymbol{\theta}}_i \tag{39}$$

Equation (39) applies the realization-wide fractional plate-scale factor $1+ s_p$ to the already catalog-perturbed position $\tilde{\theta}_i$. Thus $s_p>0$ expands the truth field radially, $s_p <0$ contracts it, and the induced angular displacement increases with the star's distance from the field center. The estimator is not given $s_p$.

The estimator retains the nominal geometry because the estimator is intentionally not told about the CAT or SCALE perturbations:

$$\boldsymbol{\theta}_i^{\text{model}} = \boldsymbol{\theta}_i^0 \tag{40}$$

Thus, the difference between truth and model coordinates represents the hidden geometric model error that the estimator must unknowingly attempt to accommodate using its states:

$$\boldsymbol{\theta}_i^{\text{true}} \neq \boldsymbol{\theta}_i^{\text{model}} \tag{41}$$

A frame-coherent radial perturbation is additionally modeled as

$$\delta r_k \sim \mathcal{N}(0, \sigma_r^2) \tag{42}$$

with stellar radial displacement magnitude in a frame while allowing the displacement direction to follow that star's own Sun-to-star radial direction:

$$\Delta\boldsymbol{\theta}_{i,k}^r = \delta r_k \hat{\mathbf{r}}_i \tag{43}$$

The synthetic truth measurement becomes

$$\mathbf{z}_{i,k}^{\text{true}} = \boldsymbol{\theta}_i^{\text{true}} + \Delta\boldsymbol{\theta}_i^{\text{GR,true}}(\gamma_0) + \delta\boldsymbol{\theta}_k^{\text{true}} + \delta\psi_k^{\text{true}}\mathbf{J}\boldsymbol{\theta}_i^{\text{true}} + \delta r_k \hat{\mathbf{r}}_i^{\text{true}} + \boldsymbol{\epsilon}_{i,k} \tag{44}$$

Equation (44) is the complete truth-generation model used in the mismatch experiments. The first term $\theta_i^{\text{true}}$ is the actual perturbed star position; $\Delta\theta_i^{\text{GR,true}}(\gamma_0)$ is the gravitational deflection evaluated at the true PPN value $\gamma_0$ and the truth geometry; $\delta\theta_k^{\text{true}}$ is the two-component frame LOS translation; $\delta\psi_k^{\text{true}}\, J\theta_i^{\text{true}}$ is the roll-induced displacement computed from the truth geometry; $\delta r_k\, \hat{r}_i^{\text{true}}$ is the optional frame-coherent RAD displacement; and $\varepsilon_{i,k}$ is random angular measurement noise. CAT and SCALE enter through $\theta_i^{\text{true}}$ whereas RAD enters through the explicit $\delta r_k\, \hat{r}_i^{\text{true}}$ term.

The estimator predicts

$$\hat{\mathbf{z}}_{i,k} = \boldsymbol{\theta}_i^0 + \Delta\boldsymbol{\theta}_i^{\text{GR,model}}(\gamma) + \delta\boldsymbol{\theta}_k + \delta\psi_k\mathbf{J}\boldsymbol{\theta}_i^0 \tag{45}$$

Equation (45) is intentionally simpler than Eq. (44) because it represents what the estimator believes. It uses the nominal catalog geometry $\theta_i^0$, the model gravitational deflection $\Delta\theta_i^{\text{GR,model}}(\gamma)$, and only the estimated LOS and roll states. No CAT, SCALE, or RAD term is included. The difference between Eqs. (44) and (45) is therefore the deliberate truth-versus-model mismatch tested in the robustness study.

Thus, the natural residual is defined as:

$$\mathbf{r}_{i,k} = \mathbf{z}_{i,k}^{\text{true}} - \hat{\mathbf{z}}_{i,k} \tag{46}$$

In Eq. (46), $r_{i,k}$ is the two-component post-model residual for star i in frame k. It contains measurement noise plus any physical displacement that the nominal estimator cannot reproduce. Under the matched Case-A model, its statistics are consistent with the assumed covariance; under hidden CAT, SCALE, or RAD mismatch, coherent structure remains in this residual.

Equations (39)-(46) are critical to the robustness analysis because the estimator is not supplied with the perturbations used to generate truth.

Figure 2 provides the spatial interpretation of these three hidden mismatch classes before they enter the estimator. CAT produces star-specific two-dimensional offsets with no common field direction; SCALE produces a coherent expansion or contraction whose displacement grows with distance from the field center; and RAD produces a common-magnitude displacement directed along each star's local solar-radial unit vector. The schematic therefore anticipates why the three perturbations need not couple to γ in the same way: only SCALE combines a field-wide radial pattern with an amplitude that systematically varies across the stellar field. This distinction motivates the source-isolation experiments of Section 3.3.

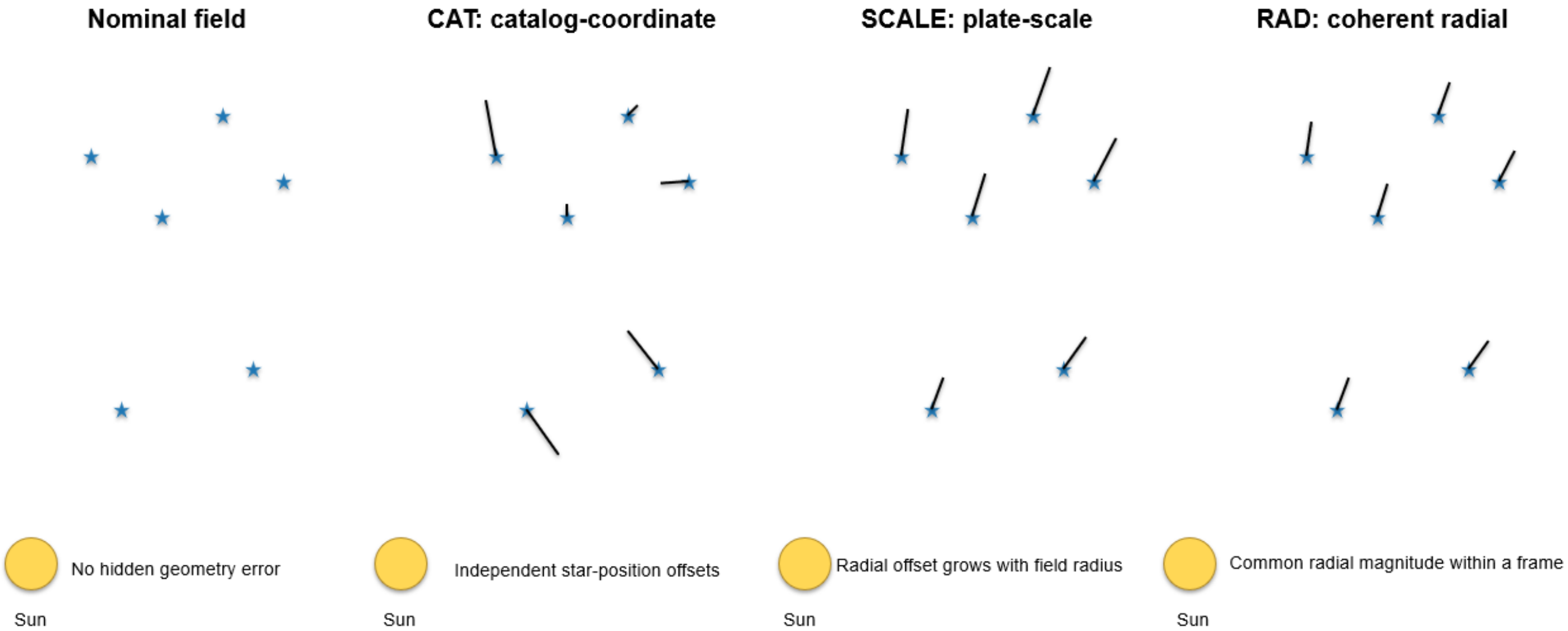

**Fig. 2. Conceptual distinction among the three hidden astrometric mismatch sources. CAT perturbs individual catalog coordinates with independent two-dimensional offsets. SCALE coherently expands or contracts the stellar field so displacement magnitude increases with field radius. RAD applies a frame-coherent radial displacement of common scalar magnitude along each star's local Sun-to-star radial direction. Arrow lengths are schematic and not to scale.**

### 2.9 Linearized Weighted Least-Squares Estimator

Estimator interpretation. Weighted least squares (WLS) chooses the state correction that minimizes residuals after each measurement is normalized by its expected uncertainty. The vector g contains the sensitivity of every star coordinate to gamma, whereas matrix A contains the three frame-local navigation signatures: x translation, y translation, and roll. The implementation does not invert the full 121-state normal matrix directly. Instead, it uses the navigation-orthogonal projector defined explicitly in Eq. (59), accumulates the reduced scalar information S and right-hand side B over all frames, and updates gamma with deltaGamma=B/S. Navigation is then re-estimated for each frame. This is implemented in estimateJointGammaNavigation() through variables g, A, AtAinv, projG, rPerp, S, B, deltaGamma, and nHat.

The complete sequence contains

$$N_z = 2N_s N_f = 2(250)(40) = 20{,}000 \tag{47}$$

The standard first-order linearization gives

$$\mathbf{r} \simeq \mathbf{H}\,\delta\mathbf{x} + \boldsymbol{\epsilon} \tag{48}$$

In Eq. (48), r is the stacked residual vector formed from all measured tangent-plane coordinates, $\mathbf{H}$ is the linearized measurement Jacobian, $\delta\mathbf{x}$ is the state correction being solved for, and $\boldsymbol{\epsilon}$ is the stacked measurement-noise vector. This relation is the local linear model from which the WLS normal equations are obtained. The relativistic Jacobian block is

$$\frac{\partial \mathbf{z}_{i,k}}{\partial \gamma} = \mathbf{g}_i \tag{49}$$

where $\mathbf{g}_i$is the 2D sensitivity of the predicted astrometric measurement of star i to the global relativistic parameter, $\gamma$.

The transverse LOS blocks are

$$\frac{\partial \mathbf{z}_{i,k}}{\partial \delta\theta_{x,k}} = \begin{bmatrix}1\\0\end{bmatrix}, \qquad \frac{\partial \mathbf{z}_{i,k}}{\partial \delta\theta_{y,k}} = \begin{bmatrix}0\\1\end{bmatrix} \tag{50}$$

Equation (50) contains two Jacobian columns. A change in $\delta\theta_{x,k}$ moves every star in frame k by the same amount in the local x-direction and therefore has derivative $[1,0]^T$; a change in $\delta\theta_{y,k}$ produces the corresponding common y-displacement and therefore has derivative $[0,1]^T$. These two columns form the translational part of the frame-local navigation matrix $\mathbf{A}_k$. The roll block is:

$$\frac{\partial \mathbf{z}_{i,k}}{\partial \delta\psi_k} = \mathbf{J}\boldsymbol{\theta}_i^0 \tag{51}$$

where a small roll produces a tangential displacement of the star so that its x-change depends on $-\theta_{y,i}$ and its y-change depends on $+\theta_{x,i}$. The three frame-local navigation signatures can now be collected explicitly into a per-star navigation Jacobian. For star i in frame k,

$$\mathbf{A}_{i,k} = \begin{bmatrix}1 & 0 & -\theta_{y,i}\\ 0 & 1 & \theta_{x,i}\end{bmatrix} \tag{52}$$

In Eq. (52), $A_{i,k}$ is the 2×3 navigation Jacobian for a single star. Its first column is the sensitivity to the frame-local x-LOS translation $\delta\theta_{x,k}$, its second column is the sensitivity to the y-LOS translation $\delta\theta_{\gamma,k}$, and its third column is the roll sensitivity. The roll column $[-\theta_{\gamma,i}, \theta_{x,i}]^T$ is tangential to the star position in the local tangent plane, consistent with the small-angle rotation operator in Eqs. (25)-(26).

$$\mathbf{A}_k = \begin{bmatrix}\mathbf{A}_{1,k}\\ \mathbf{A}_{2,k}\\ \vdots\\ \mathbf{A}_{N_s,k}\end{bmatrix} \in \mathbb{R}^{2N_s\times 3} \tag{53}$$

Equation (53) stacks the $N_s$ per-star Jacobians to form the complete frame navigation design matrix $A_k \in \mathbb{R}$^($2N_s$×3). Column 1 represents common x-LOS translation, column 2 represents common y-LOS translation, and column 3 represents position-dependent roll. This is the matrix stored as A in estimateJointGammaNavigation(); the subsequent products $A_k^T A_k$, $(A_k^T A_k)^{-1}$, and $A_k(A_k^T A_k)^{-1}A_k^T$ therefore have a direct physical interpretation as the frame-local navigation normal matrix, its inverse, and the projector onto the navigation subspace, respectively. This gives a very intuitive structure: when the sensitivities of all $N_s$ stars in frame k are stacked together $\mathbf{g}_k$ tell us how $\gamma$ moves all stars versus $\mathbf{A}_k$ tells us how navigation moves all stars.

The WLS objective is [15-17]

$$J = (\mathbf{r} - \mathbf{H}\delta\mathbf{x})^T \mathbf{R}^{-1} (\mathbf{r} - \mathbf{H}\delta\mathbf{x}) \tag{54}$$

Equation (54) is the weighted least-squares cost. The vector $\mathbf{r} - \mathbf{H}\boldsymbol{\delta}\mathbf{x}$ is the residual remaining after applying a candidate state correction; $\mathbf{R}$ is the measurement covariance matrix; and $\mathbf{R}^{-1}$ weights each residual according to its expected uncertainty. The transpose $^T$ on the left converts the weighted quadratic form to the scalar cost $J$. Minimizing $J$ gives greater influence to measurements with smaller modeled variance and leads directly to the normal equations in Eq. (55).

The normal equations are

$$\mathbf{H}^T\mathbf{R}^{-1}\mathbf{H}\,\delta\hat{\mathbf{x}} = \mathbf{H}^T\mathbf{R}^{-1}\mathbf{r} \tag{55}$$

Equation (55) is obtained by differentiating the weighted least-squares cost in Eq. (54) with respect to $\delta\mathbf{x}^T$ setting the resulting gradient equal to zero, and rearranging the first-order optimality condition. In Eq. (55), $\mathbf{H}^T\mathbf{R}^{-1}\mathbf{H}$ is the normal or information matrix and $\mathbf{H}^T\mathbf{R}^{-1}\mathbf{r}$ is the weighted residual projection onto the state sensitivities. Solving this linear system produces the correction $\boldsymbol{\delta}\mathbf{x}$ without requiring the residual itself to vanish exactly. The formal WLS solution becomes:

$$\delta\hat{\mathbf{x}} = (\mathbf{H}^T\mathbf{R}^{-1}\mathbf{H})^{-1}\mathbf{H}^T\mathbf{R}^{-1}\mathbf{r} \tag{56}$$

Equation (56) writes the formal closed-form WLS solution when the normal matrix is nonsingular. In the implemented ECLIPSE-X estimator, the equivalent solution is evaluated more efficiently by eliminating the three frame-local navigation states and solving a reduced scalar equation for $\gamma$ rather than explicitly inverting the entire 121-state matrix.

The estimator is initialized at

$$\gamma^{(0)} = 0.8 \tag{57}$$

and iterates according to

$$\mathbf{x}^{(q+1)} = \mathbf{x}^{(q)} + \delta\hat{\mathbf{x}}^{(q)} \tag{58}$$

where $\mathbf{x}^{(q)}$is the current full state estimate at iteration q, $\delta\hat{\mathbf{x}}^{(q)}$ is the estimated correction from the WLS solve at iteration q, and $\mathbf{x}^{(q+1)}$ is the updated full state estimate.

To expose the reduced scalar system used by the implementation, we define for each frame k the projector onto the complement of the frame-local navigation subspace as

$$\boldsymbol{P}_{\perp,k} = \boldsymbol{I} - \boldsymbol{A}_k\left(\boldsymbol{A}_k^T\boldsymbol{A}_k\right)^{-1}\boldsymbol{A}_k^T \tag{59}$$

In Eq. (59), $\mathbf{A}_k$ is the $2N_s\times3$ navigation design matrix for frame $k$ and $\mathbf{I}$ is the $2N_s\times2N_s$ identity matrix. The product $\mathbf{A}_k(\mathbf{A}_k^T\mathbf{A}_k)^{-1}\mathbf{A}_k^T$ projects a vector onto the subspace that can be explained by x-LOS, y-LOS, and roll. Subtracting this projector from $\mathbf{I}$ gives $\mathbf{P}_{\perp,k}$, which retains only the component that cannot be explained by those three navigation modes.

The projected relativistic sensitivity and projected residual are

$$\boldsymbol{g}_{\perp,k} = \boldsymbol{P}_{\perp,k}\boldsymbol{g}_k, \quad \boldsymbol{r}_{\perp,k} = \boldsymbol{P}_{\perp,k}\boldsymbol{r}_k \tag{60}$$

Equation (60) applies the same navigation-orthogonal projector to two quantities. $\mathbf{g}_{\perp,k}$ is the part of the $\gamma$ sensitivity that cannot be reproduced by the frame-local navigation modes, whereas $\mathbf{r}_{\perp,k}$ is the part of the measured residual that remains after those navigation modes have been removed. In the MATLAB estimator these quantities are stored as projG and rPerp, respectively.

The complementary navigation-absorbed part can also be written explicitly. For frame k, the least-squares correction to the three navigation states is

$$\boldsymbol{\delta\hat{n}_k} = \left(\boldsymbol{A}_k^T\boldsymbol{A}_k\right)^{-1}\boldsymbol{A}_k^T\boldsymbol{r}_k \tag{61}$$

In Eq. (61), $\boldsymbol{\delta\hat{n}}_k$ contains the estimated frame-local x-LOS ($\delta\hat{\theta}_{x,k}$), y-LOS ($\delta\hat{\theta}_{y,k}$), and roll adjustments ($\delta\hat{\psi}_k$) which answers the question of what combination of these variables best explains the current frame residual $\boldsymbol{r_k}$. The factor $(\mathbf{A}_k^T\mathbf{A}_k)^{-1}\mathbf{A}_k^T$ converts the projection of $\mathbf{r}_k$ onto the navigation design matrix into the least-squares navigation correction for that frame.

The complete frame residual can therefore be decomposed into a navigation-absorbed component and a navigation-orthogonal component:

$$\boldsymbol{r_k} \ = \ \boldsymbol{A_k} \ \boldsymbol{\delta\hat{n}_k} \ + \ \boldsymbol{r_{\perp,k}} \tag{62}$$

Equation (62) makes the separation explicit. The first term $\mathbf{A}_k\boldsymbol{\delta\hat{n}}_k$ is the portion of $\mathbf{r}_k$ that can be explained by changing the three frame-local navigation states. The second term $\mathbf{r}_{\perp,k}$ is orthogonal to every column of $\mathbf{A}_k$ and therefore cannot be absorbed by x-LOS, y-LOS, or roll. Only $\mathbf{r}_{\perp,k}$ can contribute directly to the reduced update of the persistent $\gamma$ state.

Information from all $N_f$ frames is then accumulated into the scalar reduced normal-equation coefficient

$$S \ = \ \sum_{k=1}^{N_f} \frac{\boldsymbol{g_{\perp,k}}^T\boldsymbol{g_{\perp,k}}}{\sigma_\theta^2} \tag{63}$$

In Eq. (63), $S$ is the scalar information accumulated for the persistent $\gamma$ state after navigation has been removed. Each frame contributes $\mathbf{g}_{\perp,k}^T\mathbf{g}_{\perp,k}$, the squared Euclidean norm of the navigation-orthogonal $\gamma$ sensitivity, divided by the angular measurement variance $\sigma_\theta^2$. Summing from $k = 1$ to $N_f$ combines the independent information carried by all frames. The corresponding weighted right-hand side

$$B \ = \ \sum_{k=1}^{N_f} \frac{\boldsymbol{g_{\perp,k}}^T\boldsymbol{r_{\perp,k}}}{\sigma_\theta^2} \tag{64}$$

In Eq. (64), $B$ is the corresponding scalar right-hand side of the reduced normal equation. For each frame, the numerator $\mathbf{g}_{\perp,k}^T\mathbf{r}_{\perp,k}$ is the inner product between the navigation-orthogonal $\gamma$ sensitivity and navigation-orthogonal residual. This inner product measures how strongly the unexplained residual aligns with the remaining $\gamma$ signature; division by $\sigma_\theta^2$ supplies the WLS weighting. The persistent-state correction is

$$\Delta\gamma \ = \ \frac{B}{S} = \frac{Residual\ evidence\ for\ changing\ \gamma}{Information\ available\ about\ \gamma} \tag{65}$$

Equation (65) is the one-dimensional reduced WLS solution for the persistent state: $B$ is the weighted residual evidence for changing $\gamma$ and $S$ is the available weighted information about $\gamma$. Their ratio gives the iterative correction $\Delta\gamma$ that is added to the current $\gamma$ estimate through Eq. (58).

Equations (59)-(65) are the explicit mathematical counterparts of A, AtAinv, projG, rPerp, S, B, and deltaGamma=B/S in estimateJointGammaNavigation(). They show both parts of the frame residual explicitly: the component absorbed by the three navigation states and the orthogonal component that remains available to update the single persistent gamma parameter.

For completeness, the matrices entering the full WLS and Fisher-information formulations can be assembled explicitly from the preceding per-star and per-frame Jacobians. For frame k, define the stacked relativistic sensitivity vector

$$\boldsymbol{g}_k \ = \ \left[\boldsymbol{g}_1^T \ \boldsymbol{g}_2^T \ \cdots \ \boldsymbol{g}_{N_s}^T\right]^T \ \in \ \mathbb{R}^{2N_s} \tag{66}$$

The corresponding frame measurement Jacobian with respect to the complete state is

$$\boldsymbol{H}_k \ = \ [\boldsymbol{g}_k \ \ 0 \ \ \cdots \ \ \boldsymbol{A}_k \ \ \cdots \ \ 0] \tag{67}$$

where $A_k \in \mathbb{R}^{2N_s\times 3}$ occupies the three columns associated with the navigation state $n_k$, while the remaining frame-local navigation blocks are zero. Stacking all $N_f$ frames gives

$$\boldsymbol{H} = \left[\boldsymbol{H}_1^T \ \boldsymbol{H}_2^T \ \cdots \ \boldsymbol{H}_{N_f}^T\right]^T \in \mathbb{R}^{2N_sN_f \times (1+3N_f)} \tag{68}$$

where there are $2N_sN_f$ astrometric measurement components and $1 + 3N_f$ estimated-state components.

Because Eq. (34) assumes independent measurements with identical variance $\sigma_\theta^2$ in both tangent-plane coordinates, the covariance of the complete stacked measurement vector and its inverse are

$$\boldsymbol{R} = \sigma^2 \boldsymbol{I}_{2N_sN_f}, \quad \boldsymbol{R}^{-1} = \frac{1}{\sigma^2}\boldsymbol{I}_{2N_sN_f} \tag{69}$$

For the present configuration, $N_s = 250$ and $N_f = 40$, so $H \in \mathbb{R}^{20000\times121}$, while $R$ and $R^{-1}$ are 20000×20000 matrices. This explicit construction connects the per-star sensitivities $g_i$ and $A_{i,k}$ to the frame matrices $g_k$ and $A_k$, the complete Jacobian $H$, and the weighting matrix $R^{-1}$.

**2.10 Fisher Information and Navigation Marginalization**

Why marginalization is necessary. A small formal uncertainty on $\gamma$ is meaningful only after accounting for the fact that LOS and roll can explain part of the same image motion. The Schur complement removes the information that can be absorbed by those nuisance states and retains only the component uniquely informative about $\gamma$. In the code this is implemented by projG=g-A*(AtAinv*(A.'*g)), followed by informationGamma=K*(projG'*projG)/sigma2 and sigmaGamma=1/sqrt(informationGamma). Thus $\sigma_\gamma$ is not the uncertainty obtained by pretending spacecraft pointing is perfectly known.

For the Gaussian linearized model [15-17], also from Eq.(55):

$$\mathbf{F} = \mathbf{H}^T\mathbf{R}^{-1}\mathbf{H} \tag{70}$$

Partitioning between γ and navigation nuisance states gives

$$\mathbf{F} = \begin{bmatrix} F_{\gamma\gamma} & \mathbf{F}_{\gamma n} \\ \mathbf{F}_{n\gamma} & \mathbf{F}_{nn} \end{bmatrix} \tag{71}$$

Equation (71) partitions the Fisher information matrix into the scalar $\gamma$ block $\mathbf{F}_{\gamma\gamma}$, the cross-information blocks $\mathbf{F}_{\gamma n}$ and $\mathbf{F}_{n\gamma}$, and the navigation-information block $\mathbf{F}_{nn}$. This partition makes explicit which information is unique to $\gamma$ and which information is shared with the nuisance navigation states.

Standard Schur-complement marginalization [15,18] gives

$$\mathcal{I}_{\gamma|n} = F_{\gamma\gamma} - \mathbf{F}_{\gamma n}\mathbf{F}_{nn}^{-1}\mathbf{F}_{n\gamma} \tag{72}$$

In Eq. (72), the term $\mathbf{F}_{\gamma n}\mathbf{F}_{nn}^{-1}\mathbf{F}_{n\gamma}$ is the portion of apparent $\gamma$ information that can be explained through coupling with navigation. Subtracting it from $F_{\gamma\gamma}$ gives the Schur-complement information $I_{\gamma|n}$, i.e., the information remaining about $\gamma$ after the navigation states have been marginalized.

Once the usable information about $\gamma$ has been obtained,

$$\sigma_\gamma = \mathcal{I}_{\gamma|n}^{-1/2} \tag{73}$$

Equation (73) converts the marginalized scalar information into the formal one-standard-deviation uncertainty. Because information is inverse variance for this linear-Gaussian model, $\sigma_\gamma^2 = I_{\gamma|n}^{-1}$ and therefore $\sigma_\gamma = I_{\gamma|n}^{-1/2}$.

Thus, the reported γ uncertainty explicitly includes information loss caused by simultaneous estimation of the 120 navigation nuisance states. This is important scientifically because the resulting $\sigma_\gamma$ represents the uncertainty after accounting for the fact that the navigation states are being estimated simultaneously.

**2.11 Monte Carlo Metrics**

Interpretation of the diagnostics. The Monte Carlo mean and bias test accuracy; $s_\gamma$ measures the observed run-to-run dispersion; the mean formal $\sigma_\gamma$ is what the estimator predicts internally; $\eta_\gamma = s_\gamma/\text{mean}(\sigma_\gamma)$ tests whether those two agree; and coverage asks how often the true $\gamma$ lies inside the estimator's nominal $1\sigma$ or $2\sigma$ interval. For a statistically calibrated Gaussian estimator, $\eta_\gamma$ should be near one and coverage should be near 0.6827 and 0.9545. runMonteCarloCase() generates each realization, calls simulateSequence() and the quiet estimator, and accumulates these quantities.

For $N_{MC}$ estimates,

$$\bar{\gamma} = \frac{1}{N_{\text{MC}}} \sum_{j=1}^{N_{\text{MC}}} \hat{\gamma}_j \tag{74}$$

Equation (74) defines the Monte Carlo mean estimate. Here $N_{\text{MC}}$ is the number of realizations, $j$ indexes an individual realization, $\hat{\gamma}_j$ is the estimate from realization $j$, and $\bar{\gamma}$ is their sample mean. The bias is defined as:

$$b_\gamma = \bar{\gamma} - \gamma_0 \tag{75}$$

Equation (75) measures estimator bias by subtracting the known truth value from the Monte Carlo mean. A value near zero indicates that the ensemble is centered on the truth; a positive value indicates systematic overestimation, and a negative value indicates systematic underestimation.

The sample standard deviation (see Equation (76)), $s_\gamma$ , is

$$s_\gamma = \sqrt{\frac{1}{N_{\text{MC}}-1} \sum_{j=1}^{N_{\text{MC}}} \left(\hat{\gamma}_j - \bar{\gamma}\right)^2} \tag{76}$$

The difference $\hat{\gamma}_j$ - $\bar{\gamma}$ is the deviation of each recovered value from the Monte Carlo mean, the squared deviations are accumulated over all realizations, and division by $N_{\text{MC}}$ - 1 gives the unbiased sample-variance estimator before taking the square root. Covariance consistency is measured by

$$\eta_\gamma = \frac{s_\gamma}{\bar{\sigma}_\gamma} \tag{77}$$

Equation (77) compares the empirical Monte Carlo dispersion with the mean formal uncertainty predicted by the estimator. A covariance-consistency ratio near unity indicates agreement between observed and predicted uncertainty; values greater than unity indicate that the estimator is overconfident, whereas values below unity indicate conservative formal uncertainty.

Coverage is

$$C_m = \frac{1}{N_{\text{MC}}} \sum_{j=1}^{N_{\text{MC}}} \mathbf{1}\left(\left|\hat{\gamma}_j - \gamma_0\right| \le m\sigma_{\gamma,j}\right) \tag{78}$$

In Eq. (78), $C_m$ is the empirical coverage probability for an $m\sigma$ interval. The indicator function equals one when the true $\gamma_0$ lies within $m$ formal standard deviations $\sigma_{\gamma,j}$ of $\hat{\gamma}_j$ and zero otherwise; averaging that indicator over all Monte Carlo realizations gives the fraction of intervals that actually contain the truth.

For a Gaussian estimator,

$$C_1 \simeq 0.6827, \qquad C_2 \simeq 0.9545 \tag{79}$$

Equation (79) gives the Gaussian reference coverage values used to judge calibration: approximately 68.27% of realizations should contain the truth within the nominal 1σ interval and approximately 95.45% within the nominal 2σ interval. Each environment uses NMC = 1000 realizations.

With $N_{\mathrm{MC}}$ = 1000, the relative standard error of $s_\gamma$ is approximately $1/\sqrt{2(N_{\mathrm{MC}}-1)} \approx$ 2.2%, and the standard error of an empirical coverage probability near 0.68 is approximately 0.015. Monte Carlo statistics are reported at full working precision so that every tabulated value can be reproduced exactly from the seeded ECLIPSE-X implementation; the digits beyond those standard errors carry reproducibility information rather than statistical precision.

**2.12 Equation-to-MATLAB Traceability**

To make the computational implementation auditable, the traceability table below maps every equation in the mathematical method to the corresponding MATLAB block, variable, or operation in the software. Equation ranges are grouped only where consecutive equations are implemented by the same code statements; every equation from (2) through (86) is covered. Section 3.1 additionally states the reduced projection-and-update sequence used to connect Eqs. (54)-(73) to the nominal numerical result. Equations (87)-(141) are reported results or diagnostic identities rather than model definitions; each is produced by the same MATLAB implementation and configuration files.

| Equations | Meaning | MATLAB implementation |
|---|---|---|
| (2)-(6) | Solar angular radius, separation, impact parameter, radial unit vector | Physical constants C.Rsun/C.AU and makeSyntheticStarCatalog(): rho_rad, b/impact_m, eRx, eRy. |
| (7)-(13) | Seeded synthetic star-field sampling and tangent-plane coordinates | rng(cfg.randomSeed); makeSyntheticStarCatalog(): u, rho_Rsun, phi, rho_rad, thetaX, thetaY. |
| (14)-(18) | PPN deflection magnitude, GR limb value, vector deflection, gamma derivative | q=2*G*Msun/(c^2*b); alphaLimb_rad; catalog.gx=q.*eRx; catalog.gy=q.*eRy. |
| (19)-(26) | Frame times, sequence duration, LOS truth, roll truth and small-angle roll operator | t=(0:Nframes-1)*cadence; truth.dThetaX, truth.dThetaY, truth.dPsi; xNav/yNav in simulateSequence(). |
| (27)-(30) | Frame-local navigation state and persistent-global state dimension | est.gammaHat plus nHat(1:3,k); K=cfg.Nframes. |
| (31)-(34) | Centroid/angular noise and Gaussian measurement model | cfg.centroidSigma_pix; cfg.angularSigma_rad; obs.sigmaRad; randn terms in obs.zx/obs.zy. |
| (35) | Matched nominal measurement equation | simulateSequence(...,1.0,0,0,0): xGR/yGR, xNav/yNav, measurement noise. |

| (36)-(37) | CAT coordinate perturbation | dCatX=catalogSigma*randn; dCatY=...; thxTrue/thyTrue. |
|---|---|---|
| (38)-(41) | SCALE perturbation and hidden truth/model geometry | plateScaleError, plateScaleTruth, thxTrue/thyTrue; obs.thetaXModel/thetaYModel remain nominal. |
| (42)-(43) | RAD frame-coherent radial perturbation | radialSys=radialSysSigma*randn; xSys=radialSys.*eRxTrue; ySys=... . |
| (44)-(46) | Truth measurement, estimator prediction, unmodeled residual | simulateSequence() truth construction plus estimateJointGammaNavigation() nominal xBase/yBase. |
| (47)-(53) | Stacked measurements, Jacobian blocks, and explicit per-star/frame navigation design matrices | g and A arrays in estimateJointGammaNavigation(); Eqs. (52)-(53) explicitly construct A_i,k and stacked A_k, whose columns are x-LOS, y-LOS, and roll. |
| (54)-(65) | WLS objective, normal equations, navigation-orthogonal projection, reduced scalar accumulation, and persistent $\gamma$ update | r, A, AtAinv, projG, rPerp, S, B, deltaGamma=B/S, gammaNew; nHat and totalCost. |
| (66)-(69) | Stacked relativistic sensitivity, frame/full Jacobian assembly, and full measurement covariance/weighting matrices | g and A arrays in estimateJointGammaNavigation(); frame stacking over K=cfg.Nframes; obs.sigmaRad supplies sigma_theta so the baseline weighting is proportional to 1/sigma_theta^2. |
| (70)-(73) | Fisher information, partitioning, Schur marginalization, formal sigma_gamma | projG projection; informationGamma=K*(projG' *projG)/sigma2; sigmaGamma=1/sqrt(informationGamma). |
| (74)-(79) | MC mean, bias, empirical dispersion, consistency ratio, coverage | runMonteCarloCase(): gammaHat/sigmaHat arrays and returned meanGamma, biasGamma, sampleSigma, meanFormalSigma, coverage1/coverage2. |
| (80)-(83) | Combined environments A/B/C | cfg.caseA, cfg.caseB, cfg.caseC: sigmaScale, catalogSigma_rad, plateScaleSigma, radialSysSigma. |
| (84)-(86) | Source-isolation CAT/SCALE/RAD ablations | cfg.ablationCAT, cfg.ablationSCALE, cfg.ablationRAD; passed to runMonteCarloCase(). |

## 3. Numerical Experiments

### 3.1 Nominal Estimation Sequence

The nominal experiment is evaluated before the robustness environments are introduced. This ordering is essential because Eqs. (80)-(86) define Monte Carlo uncertainty environments and ablation amplitudes; they are not algebraic steps used to obtain the nominal estimate reported later in Eq. (87). The nominal value of $\gamma$ is instead the output of the estimator developed in Sections 2.1-2.10 and implemented by the matched-geometry call simulateSequence(...,1.0,0,0,0), followed by estimateJointGammaNavigation().

The computational chain begins with the synthetic star geometry of Eqs. (2)-(13). Equations (14)-(18) then convert each star-Sun geometry into the leading-order PPN gravitational-deflection vector and its sensitivity to $\gamma$. The 40-frame truth histories of Eqs. (19)-(26), the persistent-global/frame-local state definition of Eqs. (27)-(30), and the angular measurement model of Eqs. (31)-(35) generate the nominal observations. In this nominal experiment, CAT, SCALE, and RAD perturbations are all zero.

For each frame $k$, the linearized residual is separated into the component that can be represented by the three frame-local navigation states and the component orthogonal to that navigation subspace. The frame navigation design matrix $\mathbf{A}_k$ is defined explicitly by Eqs. (52)-(53). Its three columns span the x-LOS translation, y-LOS translation, and roll modes that navigation is allowed to absorb.

Equation (59) then defines the frame-specific navigation-orthogonal projector $\mathbf{P}_{\perp,k}$. Applying this projector in Eq. (60) gives the projected gamma sensitivity $\mathbf{g}_{\perp,k}$ and projected residual $\mathbf{r}_{\perp,k}$. These are the portions of the gamma signature and measured residual, respectively, that lie outside the three-dimensional navigation subspace.

The complementary navigation correction $\boldsymbol{\delta}\hat{\mathbf{n}}_k$ is given by Eq. (61), and Eq. (62) makes the decomposition explicit: the navigation-absorbed part is $\mathbf{A}_k\boldsymbol{\delta}\hat{\mathbf{n}}_k$, while the remaining part is $\mathbf{r}_{\perp,k}$. In estimateJointGammaNavigation(), these operations are implemented by nHat = AtAinv*(A.'*r), rPerp = r - A*(AtAinv*(A.'*r)), and projG = g - A*(AtAinv*(A.'*g)).

Across all $N_f = 40$ frames, the code accumulates the scalar weighted information $S$ and right-hand side $B$ exactly as defined in Eqs. (63) and (64). Equation (65) then gives $\Delta\gamma = B/S$, implemented directly as deltaGamma = B/S, and Eq. (58) updates the persistent parameter. Thus the nominal solution follows the explicit sequence $\mathbf{A}_k$ -> $\mathbf{P}_{\perp,k}$ -> ($\mathbf{g}_{\perp,k}$, $\mathbf{r}_{\perp,k}$) -> ($S$, $B$) -> $\Delta\gamma$.

For the deterministic nominal realization, the estimator is initialized at $\gamma(0) = 0.8$. The first iteration gives $\Delta\gamma = 0.2001198167$ and therefore $\gamma(1) = 1.0001198167$. The second correction is $1.13 \times 10^{-16}$, so the solution is numerically unchanged at the reported precision. Consequently, Eq. (87) is a numerical result of the nominal WLS estimation chain, not a consequence of Eqs. (80)-(86).

The experiment sequence used in the remainder of the paper is therefore: (i) nominal matched-model estimation to establish recoverability and the Eq. (87) solution; (ii) combined Monte Carlo environments A, B, and C to test statistical robustness; and (iii) CAT, SCALE, and RAD source-isolation ablations to identify which hidden model discrepancy drives the observed degradation.

### 3.2 Combined Uncertainty Environments

The labels A, B, and C describe complete Monte Carlo environments, not individual physical error sources. Their numerical amplitudes are prescribed simulation parameters used to progress from a matched-model control to moderate and stronger truth-versus-model mismatch. They are not quantities derived by Eqs. (80)-(83), nor are they claimed as measured performance specifications for a particular flight instrument. The baseline angular uncertainty is inherited from the nominal measurement model of Section 2.6; the CAT, SCALE, and RAD amplitudes are controlled sensitivity-study choices encoded in cfg.caseA, cfg.caseB, and cfg.caseC.

Case A retains matched astrometric geometry and the nominal one-standard-deviation angular measurement uncertainty adopted in Section 2.6. Thus Eq. (80) is a case definition rather than a new physical derivation where the nominal random angular measurement noise is present and all structured astrometric mismatches are zero.

$$\sigma_\theta = 5.000 \times 10^{-8} \text{ rad} \tag{80}$$

Case B increases the random angular measurement uncertainty by 3.5% relative to Case A and introduces moderate hidden catalog-coordinate and plate-scale perturbations. The corresponding angular noise is $\sigma_{\theta,B} = 1.035\ \sigma_{\theta,A} = 5.175 \times 10^{-8}$ rad; $\sigma_{cat,B} = 1.0 \times 10^{-8}$ rad represents a 10-nrad one-standard-deviation catalog-coordinate perturbation, and $\sigma_{p,B} = 2.0 \times 10^{-4}$ represents a 200-ppm realization-wide fractional plate-scale uncertainty. Equation (81) defines these Case-B settings.

$$\sigma_{\text{cat}} = 1.0 \times 10^{-8} \text{ rad}, \qquad \sigma_p = 2.0 \times 10^{-4} \tag{81}$$

Case C is the stronger combined stress-test environment. Random angular measurement uncertainty is increased by 4.5% relative to Case A, giving $\sigma_{\theta,C} = 1.045\ \sigma_{\theta,A} = 5.225 \times 10^{-8}$ rad. The catalog-coordinate and plate-scale perturbations are increased to $\sigma_{cat,C} = 1.5 \times 10^{-8}$ rad (15 nrad) and $\sigma_{p,C} = 3.0 \times 10^{-4}$ (300 ppm), respectively, as defined in Eq. (82). In this case, we are increasing all values with respect to the Case B-levels.

$$\sigma_{\text{cat}} = 1.5 \times 10^{-8} \text{ rad}, \qquad \sigma_p = 3.0 \times 10^{-4} \tag{82}$$

Equation (83) additionally prescribes $\sigma_r = 2.0 \times 10^{-9}$ rad (2 nrad) as the one-standard-deviation amplitude of the frame-coherent radial perturbation used in Case C (see Eqs. (42)-(43)). The plate-scale amplitude in Case C is identical to that of the SCALE ablation, $\sigma_p = 3.0 \times 10^{-4}$, so the two environments are directly comparable. This quantity is an assumed robustness-test amplitude, not a value inferred from a particular flight sensor.

$$\sigma_r = 2.0 \times 10^{-9} \text{ rad} \tag{83}$$

Here $\sigma_\theta$ denotes random centroid-derived angular measurement noise, $\sigma_{cat}$ denotes star-specific catalog-coordinate uncertainty, $\sigma_p$ denotes fractional plate-scale uncertainty, and $\sigma_r$ denotes frame-coherent radial uncertainty. The distinction is important: $\sigma_\theta$ changes random measurement precision, whereas $\sigma_{cat}$, $\sigma_p$, and $\sigma_r$ generate structured truth-versus-model discrepancies that are intentionally hidden from the estimator. The 3.5% and 4.5% noise increases and the 10/15-nrad, 200/300-ppm, and 2-nrad perturbation levels are therefore numerical experiment design choices selected to expose sensitivity and failure modes; they should not be interpreted as experimentally measured calibration statistics.

### 3.3 Source-Isolation Ablation

The purpose of ablation is causal attribution. Cases B and C fail under several simultaneous perturbations, so they cannot by themselves identify which source is responsible. CAT, SCALE, and RAD therefore repeat the Monte Carlo experiment with baseline random measurement noise while activating only one Case-C structured mismatch at a time. This design makes the resulting changes in gamma dispersion, covariance consistency, coverage, and navigation RMS directly attributable to the isolated source.

Because B and C contain multiple simultaneous perturbations, three additional 1000-realization experiments isolate the structured error sources. The CAT experiment retains baseline centroid noise and introduces only

$$\sigma_{\text{cat}} = 1.5 \times 10^{-8} \text{ rad} \tag{84}$$

The SCALE experiment introduces only

$$\sigma_p = 3.0 \times 10^{-4} \tag{85}$$

corresponding to a 300-ppm realization-wide plate-scale uncertainty. The RAD experiment introduces only

$$\sigma_r = 2.0 \times 10^{-9} \text{ rad} \tag{86}$$

Case-C amplitudes are used in the ablation so that the expanded combined environment can be decomposed without changing the magnitude of each tested structured perturbation. So Cases B/C tell you when the estimator fails or degrades under combined mismatch whereas the Eqs. (84)-(86) tell us which mismatch caused such degradation.

## 4. Results

### 4.1 Nominal Estimation

Following the nominal estimation sequence defined in Section 3.1, not the uncertainty definitions of Eqs. (80)-(86), the matched-model solution converged from $\gamma^{(0)} = 0.8$ to

$$\hat{\gamma} = 1.0001198167 \tag{87}$$

The final error was

$$\Delta\gamma = +1.198167 \times 10^{-4} \tag{88}$$

with marginalized uncertainty

$$\sigma_\gamma = 3.075104 \times 10^{-4} \tag{89}$$

Therefore,

$$\frac{\hat{\gamma} - \gamma_0}{\sigma_\gamma} = 0.3896 \tag{90}$$

Navigation recovery is quantified as the root-mean-square difference between the estimated and prescribed truth histories over the $N_f$ frames evaluated separately for the x-LOS, y-LOS, and roll components. Navigation recovery yielded:

$$\text{RMS}_{\theta_x} = 3.132074 \times 10^{-9} \text{ rad} \tag{91}$$

$$\text{RMS}_{\theta_y} = 3.295378 \times 10^{-9} \text{ rad} \tag{92}$$

$$\text{RMS}_{\psi} = 1.499130 \times 10^{-7} \text{ rad} \tag{93}$$

The second γ correction was only $1.13\times10^{-16}$, demonstrating rapid convergence for the nominal linearized problem.

The nominal behavior is visualized in Figures 3-5. Figure 3 shows that the persistent γ estimate moves from the deliberately offset initial value $\gamma^{(0)} = 0.8$ to the converged solution in the first update, while $|\Delta\gamma|$ falls essentially to zero on the second iteration. Figures 4 and 5 provide the complementary frame-local view: the recovered x- and y-LOS histories closely follow their prescribed time histories, and the recovered roll follows the imposed oscillatory roll trajectory over the complete 195-s sequence. Taken together, these figures show that the estimator is not obtaining an accurate γ estimate by suppressing the navigation dynamics; it is simultaneously resolving the persistent relativistic parameter and the independently varying frame-local nuisance states.

### 4.2 Combined Monte Carlo Environments

The definitive v3.2 results are summarized in Table 1.

*Table 1. Combined Monte Carlo results for Cases A, B, and C.*

| **Metric** | **A** | **B** | **C** |
|---|---|---|---|
| Mean $\hat{\gamma}$ | 1.000006274 | 1.025348436 | 0.904362690 |
| Bias | $+6.2744\times10^{-6}$ | $+2.5348\times10^{-2}$ | $-9.5637\times10^{-2}$ |
| Empirical $s_\gamma$ | $3.039710\times10^{-4}$ | 1.549842 | 2.224386 |
| Formal $\bar{\sigma}_\gamma$ | $3.075104\times10^{-4}$ | $3.182733\times10^{-4}$ | $3.213484\times10^{-4}$ |
| $C_1$ | 0.693 | 0.000 | 0.000 |
| $C_2$ | 0.951 | 0.000 | 0.001 |
| LOS RMS [rad] | $3.180427\times10^{-9}$ | $2.121935\times10^{-7}$ | $3.046895\times10^{-7}$ |
| Roll RMS [rad] | $1.463463\times10^{-7}$ | $2.903096\times10^{-7}$ | $3.774608\times10^{-7}$ |
| Solver failure | 0% | 0% | 0% |

Case A gives

$$\eta_{\gamma,A} = \frac{3.039710\times10^{-4}}{3.075104\times10^{-4}} = 0.9885 \tag{94}$$

which means that the actual Monte Carlo scatter is approximately equal to the uncertainty predicted by the estimator, so Case A demonstrates that when the truth and estimator models are matched, the covariance prediction is properly calibrated. The empirical and formal uncertainties therefore differ by only approximately 1.2%. Its coverage is

$$C_{1,A} = 0.693 \ \approx 0.6827, \qquad C_{2,A} = 0.951 \ \approx 0.9545 \tag{95}$$

Cases B and C behave fundamentally differently. Their empirical dispersions increase by several orders of magnitude while the formal covariance remains near its matched-model value. Both environments therefore demonstrate severe covariance inconsistency under hidden astrometric model mismatch.

Figure 6 makes this loss of statistical robustness visible in the Monte Carlo distributions themselves. The Case-A distribution remains narrow and centered close to $\gamma = 1$, consistent with the empirical dispersion and coverage reported in Table 1. In contrast, the B and C distributions broaden over order-unity ranges of $\gamma$ and are displaced relative to the matched-model distribution. The visual separation is important because the formal $\sigma_\gamma$ values change only slightly across the three environments: the estimator continues to report a small internal uncertainty even while the realized $\gamma$ estimates become widely dispersed under hidden model mismatch.

### 4.3 Source-Isolation Ablation Results

The source-isolation ablation identifies plate-scale mismatch as the dominant origin of this failure. In Table 2, the formal uncertainty, $\bar{\sigma}_\gamma$ , remains identical across the source-isolation cases because CAT, SCALE, and RAD are injected only into the truth model and are intentionally absent from the estimator state and assumed covariance; differences among the cases therefore appear in the empirical scatter and coverage rather than in the estimator-reported formal uncertainty.

*Table 2. Source-isolation ablation results.*

| **Metric** | **CAT** | **SCALE** | **RAD** |
|---|---|---|---|

| Mean $\hat{\gamma}$ | 1.000063797 | 1.038044773 | 0.999993297 |
|---|---|---|---|
| Bias | $+6.3797\times10^{-5}$ | $+3.8045\times10^{-2}$ | $-6.7034\times10^{-6}$ |
| Empirical $s_\gamma$ | $6.714457\times10^{-4}$ | 2.191498 | $3.452761\times10^{-4}$ |
| Formal $\bar{\sigma}_\gamma$ | $3.075104\times10^{-4}$ | $3.075104\times10^{-4}$ | $3.075104\times10^{-4}$ |
| $\eta_\gamma$ | 2.1835 | 7126.6 | 1.1228 |
| $C_1$ | 0.357 | 0.000 | 0.626 |
| $C_2$ | 0.615 | 0.000 | 0.928 |
| LOS RMS [rad] | $3.2990\times10^{-9}$ | $2.9800\times10^{-7}$ | $3.1597\times10^{-9}$ |
| Roll RMS [rad] | $1.5228\times10^{-7}$ | $3.6637\times10^{-7}$ | $1.4619\times10^{-7}$ |

The resulting sensitivity, $\eta_\gamma$ , hierarchy is

$$\mathrm{RAD} < \mathrm{CAT} \ll \mathrm{SCALE} \tag{96}$$

Catalog-coordinate mismatch approximately doubles the empirical $\gamma$ uncertainty relative to the formal prediction:

$$\eta_{\gamma,\mathrm{CAT}} = 2.1835 \tag{97}$$

The tested radial systematic has a smaller effect:

$$\eta_{\gamma,\mathrm{RAD}} = 1.1228 \tag{98}$$

In contrast, plate-scale mismatch alone produces

$$s_{\gamma,\mathrm{SCALE}} = 2.191498 \tag{99}$$

against a formal uncertainty of only $3.075104\times10^{-4}$. Thus,

$$\eta_{\gamma,\mathrm{SCALE}} = 7126.6 \tag{100}$$

Both coverage measures collapse to zero.

Figures 7 and 8 place the ablation results and the combined environments on common statistical diagnostics. Figure 7 shows the empirical-to-formal ratio $\eta_\gamma$ on a logarithmic scale: A remains near the ideal value of unity, RAD remains close to unity, CAT rises to 2.1835, whereas SCALE increases to 7126.6; the combined B and C environments are likewise grossly inconsistent. Figure 8 shows the corresponding coverage consequence. Case A lies close to the Gaussian $1\sigma$ and $2\sigma$ reference levels, CAT and RAD show partial degradation, but SCALE and B lose both $1\sigma$ and $2\sigma$ coverage, while C retains only 0.001 at $2\sigma$. The two figures therefore distinguish numerical dispersion from covariance calibration: the principal failure is not simply a larger error bar, but an uncertainty model that no longer represents the realized estimation error.

### 4.4 Plate-Scale Error Dominates the Combined Failure

For a small-scale error, Eq. (39) gives a first-order field-wide displacement

$$\Delta\boldsymbol{\theta}^{\mathrm{scale}} \simeq s_p\boldsymbol{\theta}^0 \tag{101}$$

The plate-scale residual is therefore coherent and solar-centered, with radial magnitude proportional to field radius. Gravitational displacement is also radial but follows

$$\Delta \boldsymbol{\theta}^{\mathrm{GR}} \propto \rho^{-1}\hat{\mathbf{r}} \tag{102}$$

Thus, the two signatures have different radial dependence but occupy related radial spatial modes. Because the present estimator contains no scale-calibration state, the scale-induced residual must be accommodated by the available $\gamma$, LOS, and roll states. The navigation results support this interpretation. CAT and RAD retain LOS RMS near the baseline value, whereas SCALE increases LOS RMS to

$$\mathrm{RMS}_{\mathrm{LOS,SCALE}} = 2.980012 \times 10^{-7}\ \mathrm{rad} \tag{103}$$

approximately two orders of magnitude above Case A. The ablation therefore identifies plate-scale mismatch -not the tested catalog or radial perturbation -as the dominant contributor to the B/C breakdown.

Figure 9 connects this statistical failure to the recovered spacecraft states. The isolated CAT and RAD cases leave LOS RMS near the matched baseline and produce only modest changes in roll RMS, whereas SCALE drives both navigation metrics upward, with LOS RMS reaching $2.9800 \times 10^{-7}$ rad and roll RMS $3.6637 \times 10^{-7}$ rad. Cases B and C exhibit the same elevated-error regime. This confirms that the unmodeled scale deformation is not confined to the $\gamma$ estimate because SCALE is absent from the estimated state, its coherent field signature is redistributed among the persistent relativistic state and the available frame-local LOS and roll states.

### 4.5 Numerical Convergence versus Statistical Validity

Every realization in A, B, C, CAT, SCALE, and RAD completed without a numerical solver failure. Nevertheless, SCALE produces $C_1 = C_2 = 0$ despite numerical convergence, with $\eta_\gamma = 7126.6$ demonstrating that the formal covariance substantially underestimates the empirical uncertainty under plate-scale mismatch. Therefore,

$$\text{numerical convergence} \not\Rightarrow \text{statistical validity} \tag{104}$$

Solver status alone cannot reveal a large coherent model discrepancy that lies outside the assumed state and covariance model.

The complete figure set reinforces this distinction. Figure 3 shows clean numerical convergence, and Figures 4-5 show successful nominal navigation recovery; however, Figures 6-9 show that the same solver can remain numerically well behaved while hidden astrometric mismatch destroys statistical calibration. Convergence is therefore a necessary computational property, not evidence by itself that the assumed state and covariance model are physically adequate.

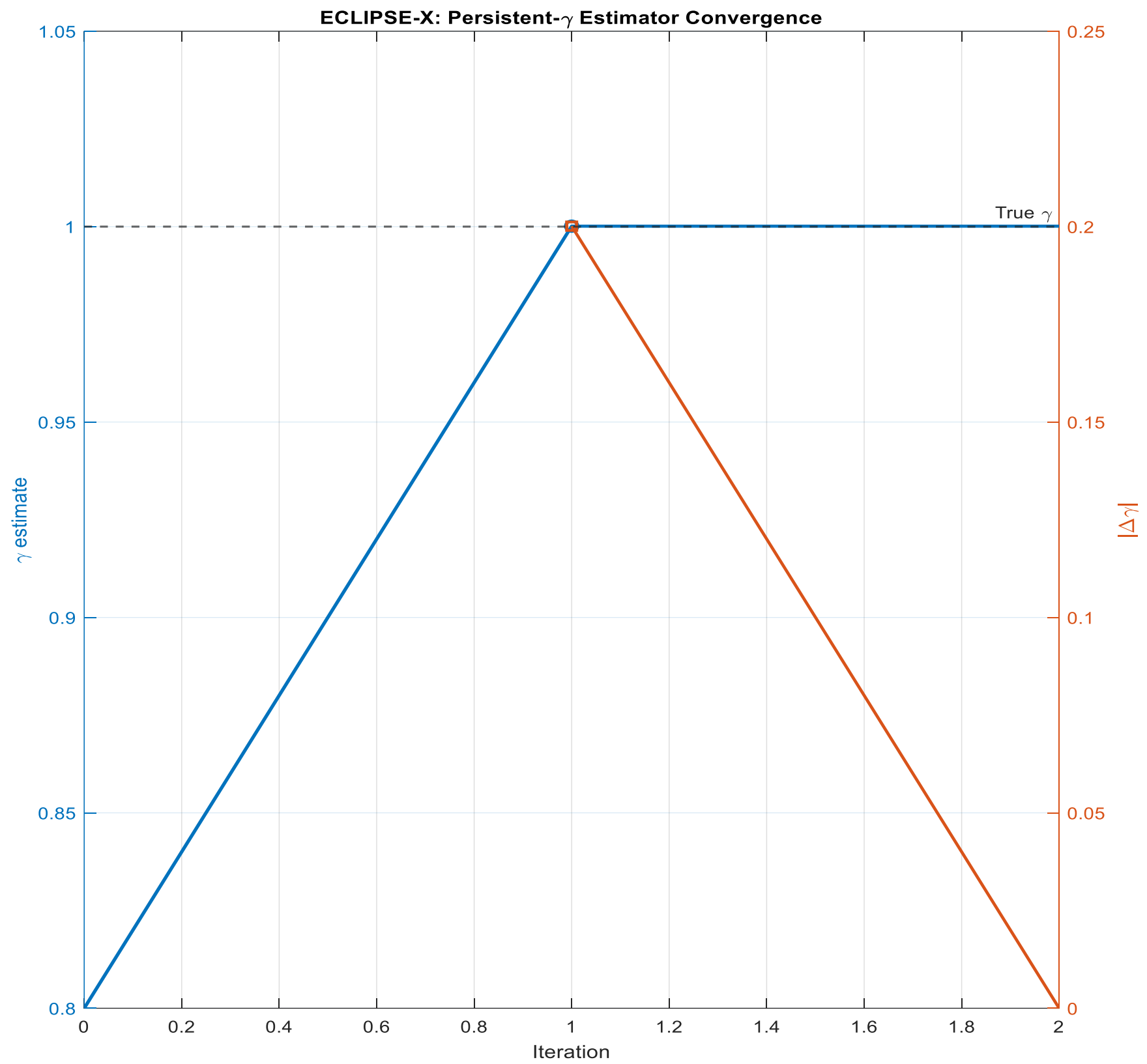

**Fig. 3. Convergence history of the persistent PPN γ estimator from the initial value $\gamma^{(0)} = 0.8$ to the converged nominal estimate; the secondary axis shows the magnitude of the iterative correction |Δγ|.**

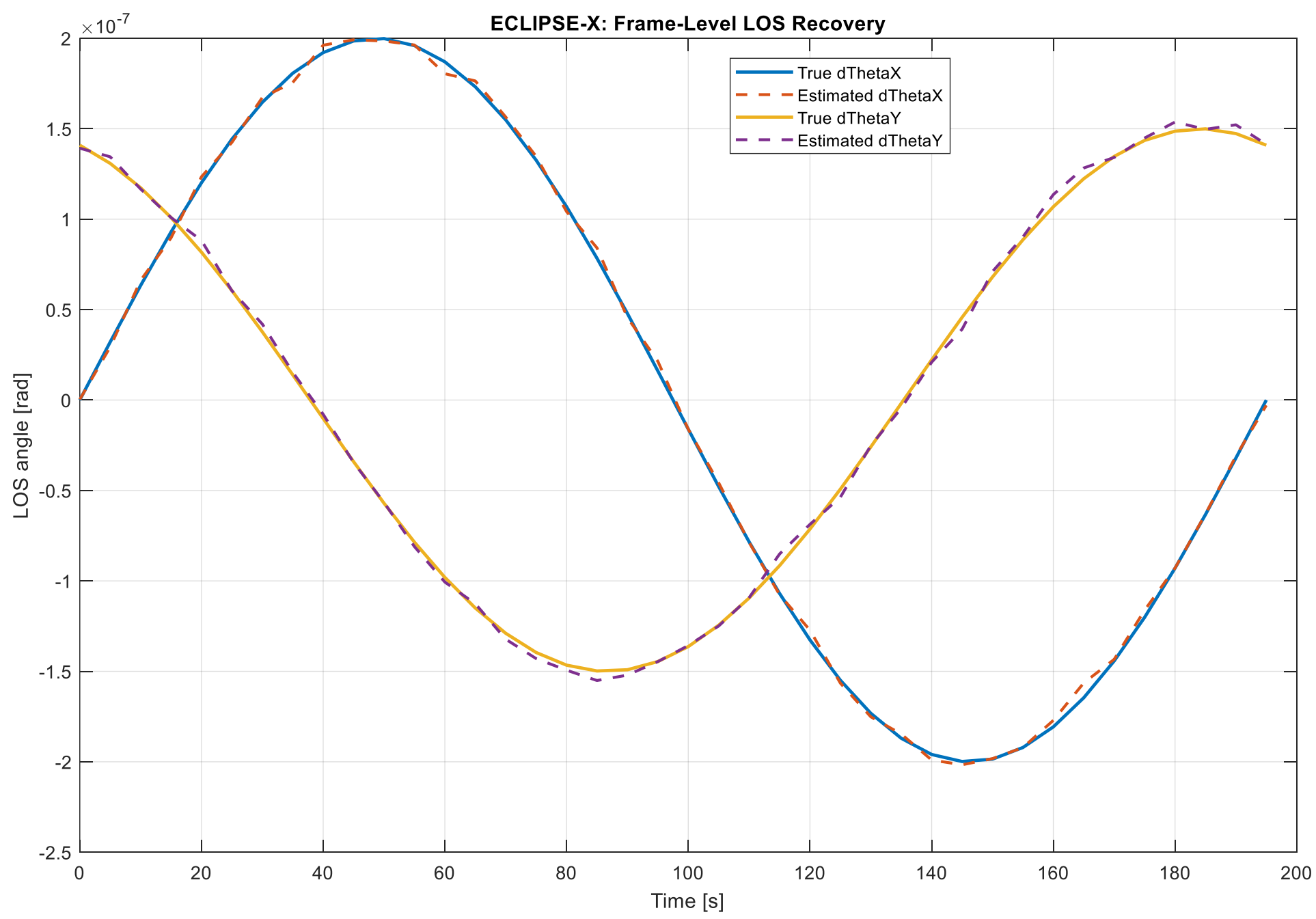

**Fig. 4. Frame-level recovery of the transverse line-of-sight states Δθx and Δθy over the 40-frame, 195-s observation sequence, comparing prescribed truth with the estimated histories.**

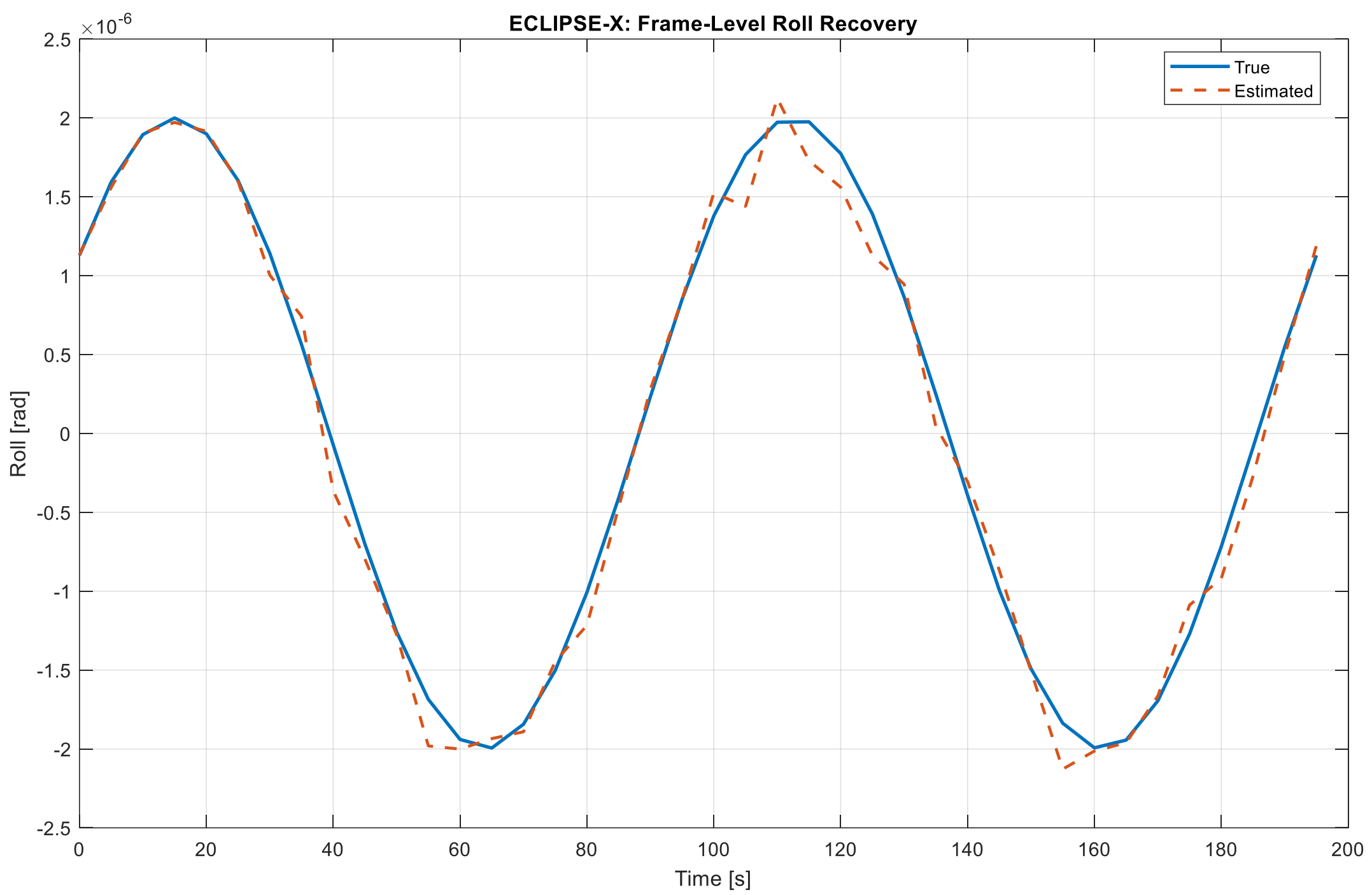

**Fig. 5. Frame-level recovery of spacecraft roll Δψ over the 40-frame, 195-s observation sequence, comparing the prescribed truth and estimated roll histories.**

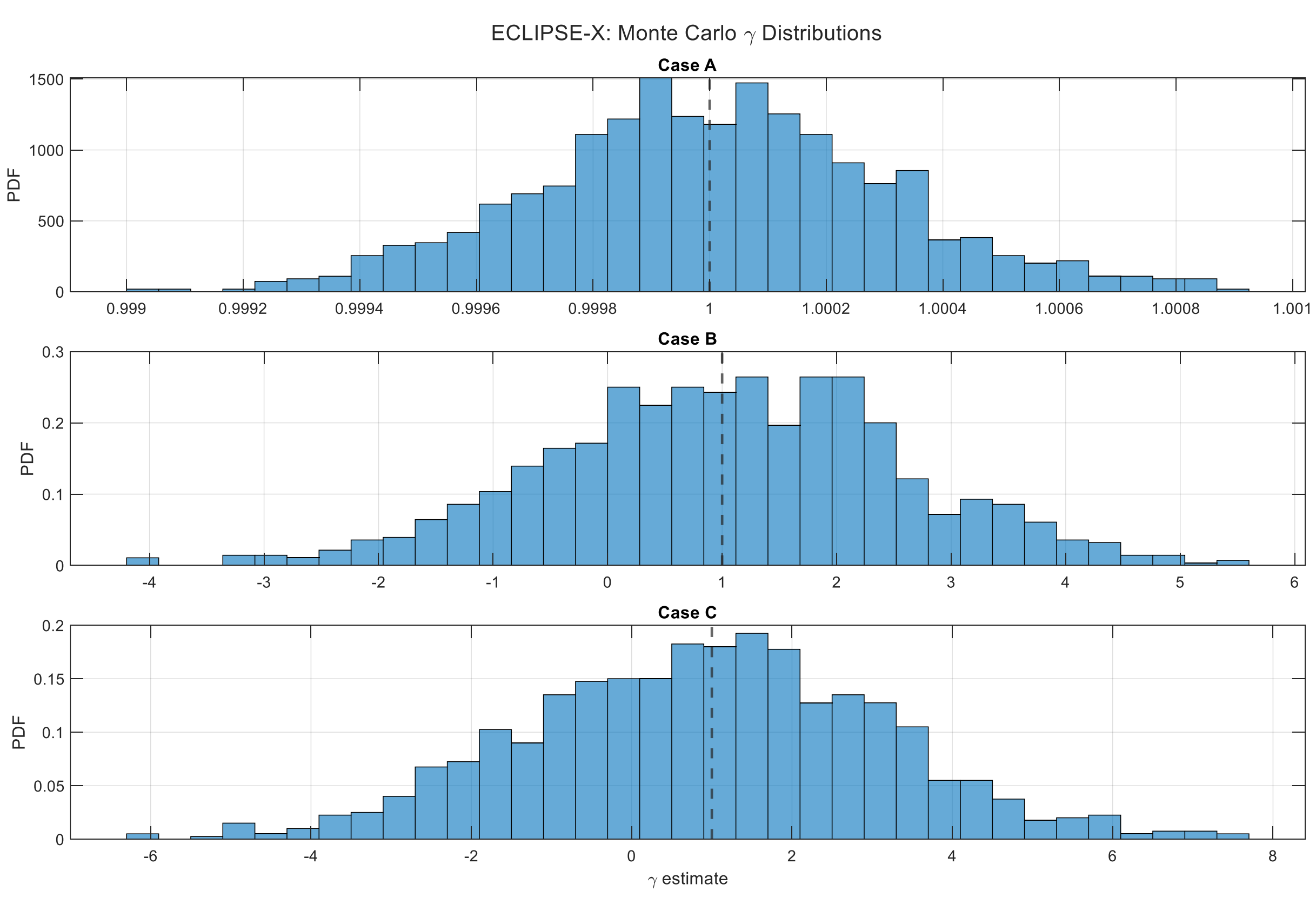

.

**Fig. 6. Monte Carlo distributions of the estimated PPN parameter γ for combined environments A, B, and C. Case A represents matched astrometric geometry, whereas Cases B and C introduce progressively larger hidden astrometric model mismatch.**

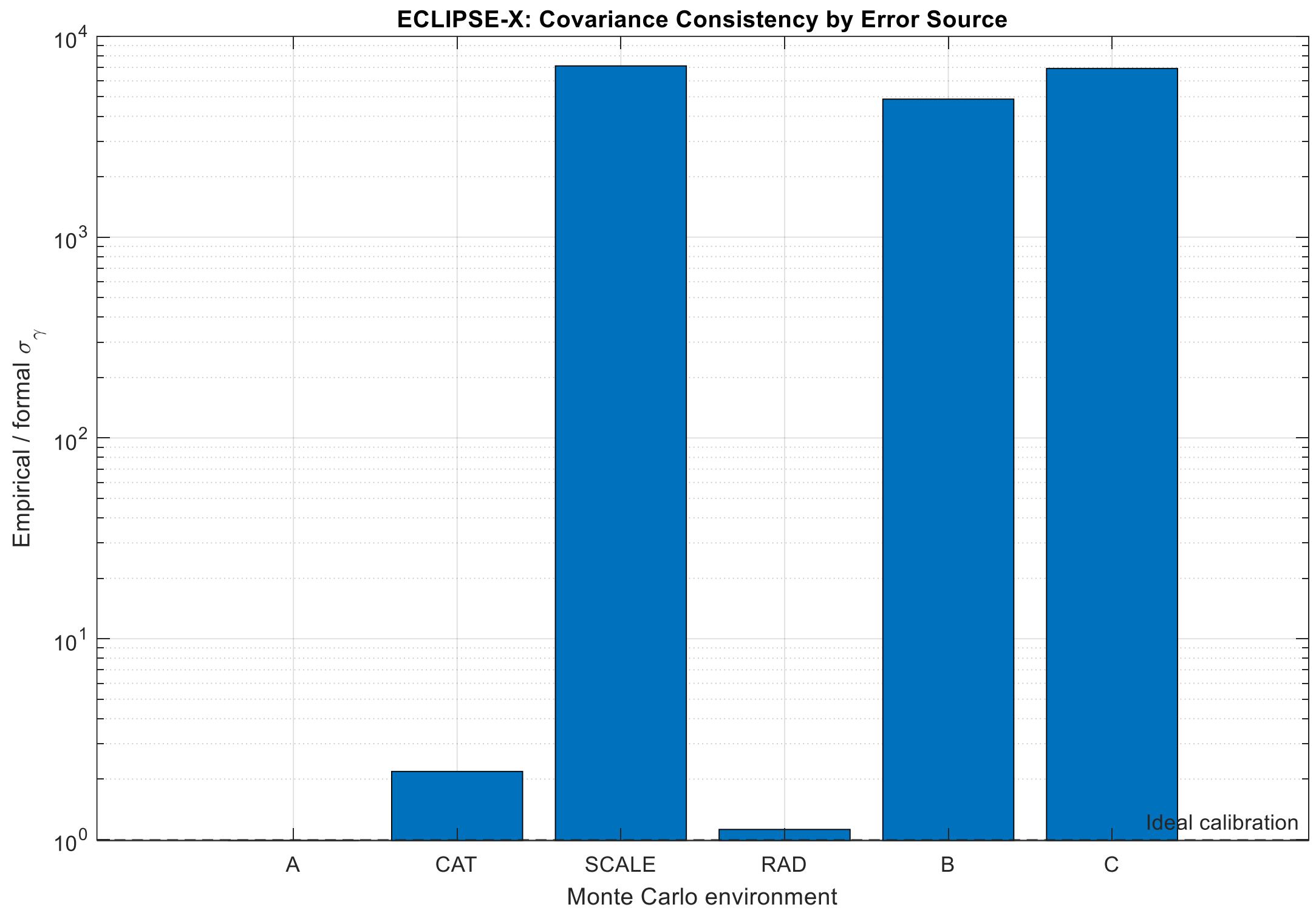

**Fig. 7. Empirical-to-formal γ uncertainty ratio ηγ for the matched baseline (A), isolated catalog-coordinate (CAT), plate-scale (SCALE), and radial (RAD) perturbations, and combined mismatch environments B and C. The ideal covariance-consistency value $\eta_\gamma = 1$ is indicated.**

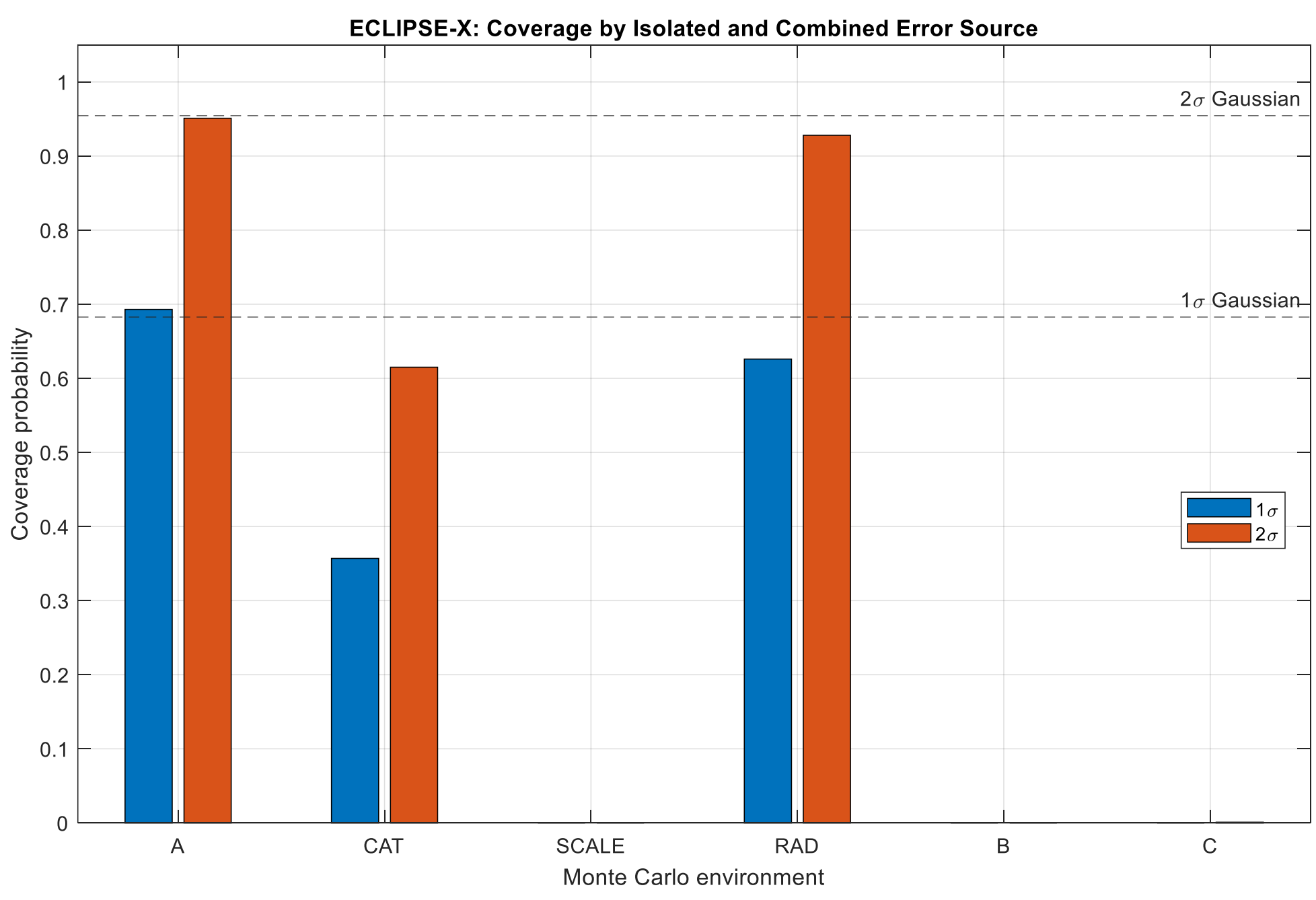

**Fig. 8. Empirical 1σ and 2σ coverage probabilities for the matched baseline, isolated error-source ablations, and combined mismatch environments. Dashed reference lines indicate the nominal Gaussian coverage probabilities.**

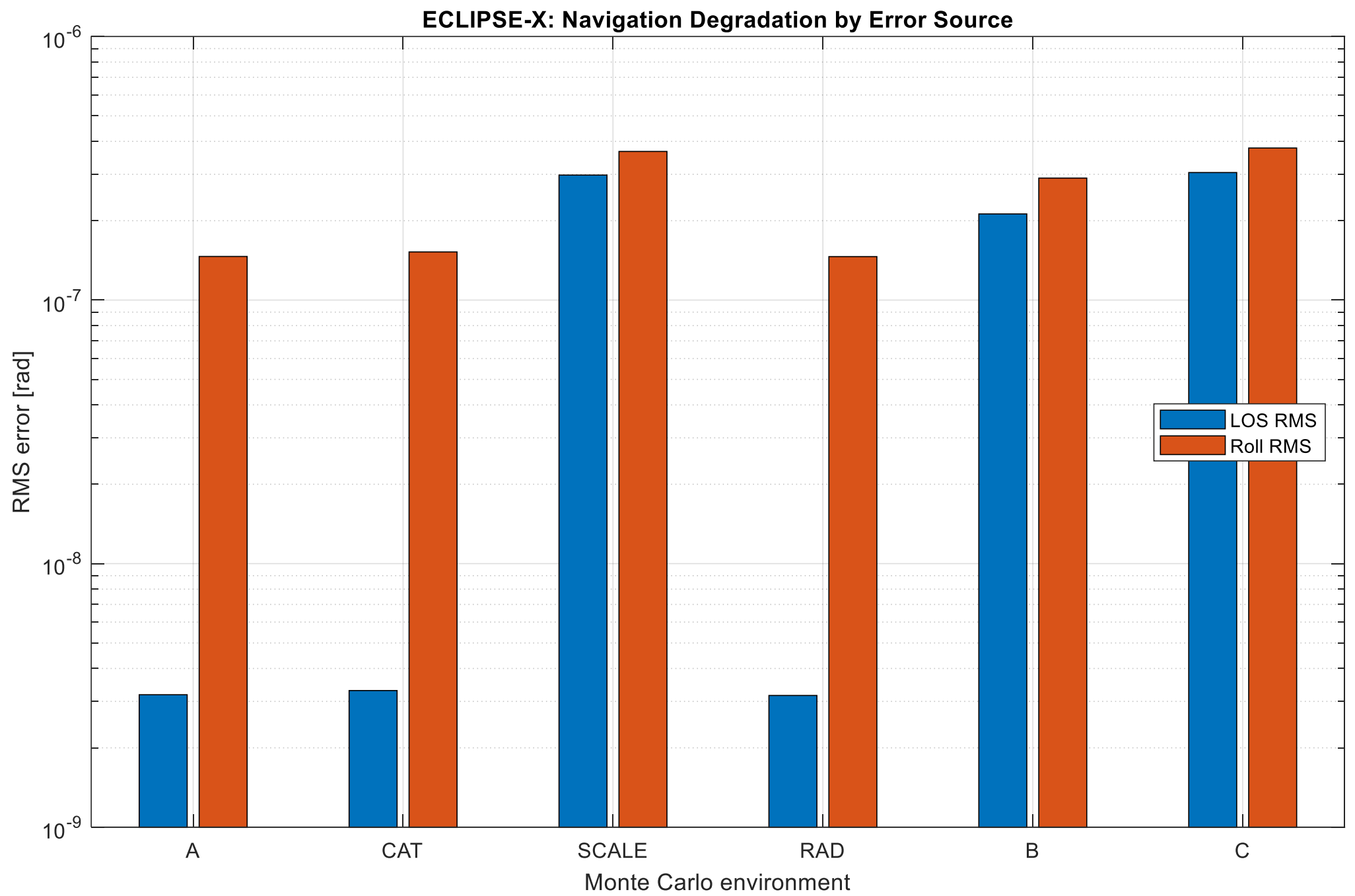

**Fig. 9. Navigation-state degradation across the matched baseline, isolated error-source ablations, and combined mismatch environments, expressed as RMS transverse line-of-sight and roll estimation errors.**

## 5. Discussion

### 5.1 Validation of the Persistent-Global/Frame-Local Architecture

The nominal solution and Case A demonstrate that γ, transverse LOS, and roll can be recovered simultaneously under the modeled conditions. The architecture benefits from complementary spatial and temporal separation. The first-order relativistic signature follows

$$\Delta\boldsymbol{\theta}^{\mathrm{GR}} \propto \rho^{-1}\hat{\mathbf{r}} \tag{105}$$

where the stars close to the solar limb experience the strongest GR deflection, while further stars from the Sun experience progressively smaller deflection; whereas transverse LOS error is approximately translational and roll is tangential and position dependent.

Temporally, the same γ contributes to all 40 frames while the navigation states are independently estimated at each epoch. The matched-model Monte Carlo ensemble gives

$$\eta_{\gamma,A} = \frac{s_{\gamma,A}}{\bar{\sigma}_{\gamma,A}} = 0.9885 \tag{106}$$

which independently validates the Fisher-derived marginalized covariance under matched assumptions. The corresponding coverage values are close to the Gaussian reference probabilities.

This matched-model interpretation is also supported visually by Figures 3-5: rapid persistent-state convergence occurs simultaneously with close tracking of the prescribed LOS and roll histories. Figure 6 then provides the ensemble counterpart, with Case A forming the narrow distribution around the GR value before hidden astrometric mismatch is introduced.

### 5.2 Catalog Error Is Important but Not Dominant

The CAT ablation introduces only the Case-C catalog-coordinate perturbation while retaining the baseline centroid-noise level. Its empirical $\gamma$ dispersion is

$$s_{\gamma,\mathrm{CAT}} = 6.714457 \times 10^{-4} \tag{107}$$

compared with the formal prediction $3.075104 \times 10^{-4}$, giving

$$\eta_{\gamma,\mathrm{CAT}} = 2.1835 \tag{108}$$

and coverage

$$C_{1,\mathrm{CAT}} = 0.357, \qquad C_{2,\mathrm{CAT}} = 0.615 \tag{109}$$

Thus catalog-coordinate uncertainty cannot be ignored at the precision examined here. Accordingly, only 35.7% of the CAT realizations contain the true $\gamma^{(0)}$ within their nominal $1\sigma$ instead of approximately 68.27%. Similarly, 61.5% contain truth within $2\sigma$, instead of approximately 95%. However, LOS and roll recovery remain close to the baseline solution, and the resulting degradation is orders of magnitude smaller than the plate-scale case.

The intermediate role of CAT is evident in Figures 7-9. Its $\eta_\gamma$ exceeds the ideal value and its coverage falls below the Gaussian references, yet its navigation RMS remains close to the matched baseline. CAT therefore degrades statistical calibration without producing the system-wide state contamination observed for SCALE.

### 5.3 Tested Radial Systematic Produces Moderate Degradation

The RAD experiment introduces only the frame-coherent radial perturbation defined in the numerical experiment. The empirical-to-formal uncertainty ratio becomes

$$\eta_{\gamma,\mathrm{RAD}} = 1.1228 \tag{110}$$

with coverage

$$C_{1,\mathrm{RAD}} = 0.626, \qquad C_{2,\mathrm{RAD}} = 0.928 \tag{111}$$

The tested $2.0 \times 10^{-9}$-rad perturbation therefore produces measurable covariance underestimation but does not destroy the relativistic solution. This result applies only to the specific magnitude, spatial form, and frame-to-frame behavior evaluated here.

Figures 7 and 8 show this moderate effect directly: RAD remains much closer to ideal covariance consistency and Gaussian coverage than CAT, SCALE, B, or C. Figure 9 likewise shows that the RAD navigation errors remain essentially at the nominal scale. This behavior is specific to the tested 2 nrad frame-coherent radial perturbation and should not be extrapolated to other radial-systematic structures or amplitudes.

### 5.4 Plate Scale Is the Dominant Observability Requirement

The SCALE ablation introduces only the realization-wide plate-scale mismatch at the Case-C level,

$$\sigma_p = 3.0 \times 10^{-4} = 300\ \mathrm{ppm} \tag{112}$$

while retaining baseline centroid noise. The resulting empirical $\gamma$ dispersion is

$$s_{\gamma,\mathrm{SCALE}} = 2.191498 \tag{113}$$

so even a small 0.03% error becomes important across a wide angular field. The covariance-consistency ratio becomes

$$\eta_{\gamma,\mathrm{SCALE}} = 7126.6 \tag{114}$$

with zero 1σ and 2σ coverage. The same case increases LOS RMS to 2.980012 × $10^{-7}$ rad and roll RMS to 3.663714 × $10^{-7}$ rad. The perturbation therefore contaminates the entire state solution rather than only γ.
This state-wide contamination is the central visual result of Figures 7-9. Figure 7 identifies SCALE as the dominant covariance-consistency failure by several orders of magnitude, Fig. 8 shows the accompanying collapse of interval coverage, and Fig. 9 shows that the same mismatch simultaneously degrades LOS and roll recovery. The three diagnostics therefore support a single physical interpretation: an unestimated plate-scale degree of freedom projects coherently into both the scientific and navigation portions of the solution.
For small plate-scale error, the induced field displacement is approximately

$$\Delta\boldsymbol{\theta}^{\mathrm{scale}} \simeq s_p\boldsymbol{\theta}^0 \tag{115}$$

The radial growth of the SCALE signature follows directly from the tangent-plane geometry. Because the nominal stellar coordinate can be written as a radial magnitude multiplied by its local solar-radial unit vector,

$$\theta_i{}^0 = \rho_i\hat{r}_i \tag{116}$$

Substituting into Eq. (115) yields that the plate-scale error is also radial with respect to the solar center:

$$\Delta\theta_i{}^{scale} \simeq s_p\rho_i\hat{r}_i \tag{117}$$

and therefore the magnitude of the plate-scale displacement is

$$\left\|\Delta\theta_i{}^{scale}\right\| = \left|s_p\right|\rho_i \tag{118}$$

Equations (117)-(118) make the physical behavior explicit: SCALE produces a coherent radial expansion or contraction of the stellar field, and the absolute angular displacement grows linearly with distance $\rho_i$ from the solar center. This is different from transverse LOS error, which is approximately a uniform translation of the field, and from roll error, which produces a tangential displacement.

The important comparison with solar gravitational deflection is that both SCALE and GR are radial, but their radial dependences are opposite. To leading order,

$$\Delta\theta^{\mathrm{GR}} \propto \frac{1}{\rho}, \qquad \Delta\theta^{\mathrm{scale}} \propto \rho \tag{119}$$

Consequently, the two signatures are not mathematically identical. GR is strongest for stars nearest the solar limb and decreases approximately as $1/\rho_i$, whereas a fractional plate-scale error produces a displacement that increases as $\rho_i$. The broad 1.22-8 $R_\odot$ radial distribution used in the controlled field therefore provides spatial diversity that helps distinguish these signatures; nevertheless, if scale error is absent from the estimated state, its coherent residual must still project onto the available $\gamma$, LOS, and roll degrees of freedom.

The magnitude of the tested 300-ppm SCALE perturbation can be understood directly. Using the approximate apparent solar angular radius $\rho_\odot \approx 4.65 \times 10^{-3}$ rad, a star near the outer edge of the simulated field has

$$\rho \approx 8\rho_\odot \approx 8(4.65 \times 10^{-3}) = 3.72 \times 10^{-2} rad \tag{120}$$

For a representative $1\sigma$ scale-error magnitude $\left|s_p\right| = \sigma_p = 3.0 \times 10^{-4}$, the corresponding scale displacement at that radius is therefore:

$$\left|\Delta\theta^{\text{scale}}\right| \approx (3.0 \times 10^{-4})(3.72 \times 10^{-2}) = 1.12 \times 10^{-5} rad \approx 2.3\, arcsec \quad (121)$$

By comparison, the baseline random angular measurement uncertainty is $\sigma_\theta$ = 5.0 × $10^{-8}$ rad ≈ 0.0103 arcsec. Thus, at the outer field radius, the representative 300-ppm scale displacement is approximately

$$\frac{|\Delta\theta^{\text{scale}}|}{\sigma_\theta} \approx 224 \quad (122)$$

times the assumed one-standard-deviation random angular measurement noise. This calculation explains why a fractional calibration error that appears numerically small can dominate a high-precision astrometric estimator. It also explains the severe SCALE covariance inconsistency observed in the Monte Carlo results: the formal estimator covariance remains small because plate scale is not represented as an estimated state or modeled uncertainty, while the truth data contain a coherent field deformation much larger than the assumed random noise over much of the field.

The plate-scale displacement is coherent, solar-centered, and radial, with amplitude increasing with field radius. The GR displacement is also radial but follows the opposing first-order radial dependence

$$\Delta\boldsymbol{\theta}^{\text{GR}} \propto \rho^{-1}\hat{\mathbf{r}} \quad (123)$$

so the two spatial modes are not identical. However, because plate scale is absent from the estimated state, the unmodeled coherent scale residual must project onto the available γ, LOS, and roll degrees of freedom. The ablation establishes the hierarchy

$$\text{RAD} < \text{CAT} \ll \text{SCALE} \quad (124)$$

for the perturbation magnitudes tested here. Plate-scale knowledge must therefore be treated as an observability-critical calibration quantity for this estimator architecture.

### 5.5 Formal Covariance Is Conditional on Model Adequacy

The formal covariance is derived from the assumed linearized information matrix

$$\mathbf{F} = \mathbf{H}^T\mathbf{R}^{-1}\mathbf{H} \quad (125)$$

where the formal covariance is determined by the state sensitivities represented in **H** and the measurement uncertainties represented in **R**; therefore, error sources absent from both the estimator state model and assumed measurement covariance do not contribute to the predicted uncertainty. When plate-scale error is absent from those assumptions, the Fisher matrix contains no term representing that source of uncertainty. Consequently, the formal γ uncertainty remains approximately

$$\bar{\sigma}_{\gamma,\text{SCALE}} = 3.075104 \times 10^{-4} \quad (126)$$

while the empirical distribution has

$$s_{\gamma,\text{SCALE}} = 2.191498 \quad (127)$$

The discrepancy is therefore a model-adequacy failure, not a contradiction of Fisher-information theory. It illustrates why covariance validation must include perturbations that are not represented in the estimator's nominal model.

### 5.6 Numerical Convergence Does Not Imply Statistical Validity

All realizations in A, B, C, CAT, SCALE, and RAD completed without a numerical solver failure. Yet SCALE produced zero 1σ and 2σ coverage and an empirical-to-formal uncertainty ratio exceeding seven thousand. Hence

$$\text{numerical convergence } \not\Rightarrow \text{ statistical validity} \tag{128}$$

A solver can meet its iteration and convergence criteria while producing a scientifically invalid confidence interval if the data contain coherent model discrepancies absent from the estimator. Eq. (128) states that successful numerical convergence only shows that the estimator solved its assumed model; it does not prove that its estimated state or covariance is statistically consistent with physical truth.

### 5.7 Joint Plate-Scale Estimation Restores Covariance Consistency

For spacecraft navigation, the operational result is that solver convergence alone is not a sufficient validity test. In the SCALE ablation, a realization-wide calibration error degraded the recovered LOS and roll states together with $\gamma$ even though every numerical realization completed successfully. The implication extends beyond this specific relativistic experiment: whenever a coherent optical-calibration mode has appreciable projection onto the measurement signatures of the estimated states, that mode must be independently constrained, represented in the covariance model, or estimated jointly. Otherwise, a navigator may obtain a precise-looking state estimate and formal covariance that are statistically inconsistent with the true error. The ECLIPSE-X results therefore convert plate-scale knowledge from a secondary camera-calibration detail into an observability-critical requirement for this class of precision stellar-navigation estimator.

The ablation identifies the estimator extension directly. Plate scale is introduced as a persistent calibration state together with $\gamma$ and the frame-local navigation states:

$$\mathbf{x}_{\text{aug}} = \begin{bmatrix} \gamma & s_p & \mathbf{n}_1^T & \mathbf{n}_2^T & \cdots & \mathbf{n}_{N_f}^T \end{bmatrix}^T \tag{129}$$

The augmented formulation is potentially identifiable because $\gamma$ and $s_p$ generate different radial basis functions whereas the navigation state $n_k$ remains frame dependent. The augmented formulation therefore contains 122 states rather than 121.

To first order, their measurement sensitivities scale as

$$\frac{\partial \theta_i}{\partial \gamma} \propto \frac{1}{\rho_i}\hat{r}_i, \qquad \frac{\partial \theta_i}{\partial s_p} = \rho_i \hat{r}_i - (1+\gamma) q_i \hat{r}_i \tag{130}$$

The second term in the plate-scale sensitivity is not optional. It is collinear with the $\gamma$ column, so omitting it does not change the column space of the augmented design matrix but reparametrizes it: the recovered coefficient becomes $\hat{\gamma} - 2s_p$ rather than $\hat{\gamma}$, and the Monte Carlo dispersion inflates from $\sigma_\gamma$ to $(\sigma_\gamma{}^2 + 4\sigma_p{}^2)^{1/2}$ with no accompanying change in the formal covariance. The naive column $\rho_i \hat{r}_i$ therefore reproduces, in a different guise, precisely the covariance inconsistency the augmentation is intended to remove.

Whether the augmented system remains well conditioned at the radial sampling used here does not follow from Eq. (130) alone, since the separability of the $\rho^{-1}$ and $\rho$ columns depends on the realized distribution of $q_i$. It is resolved here directly. Implementing Eq. (129) as a 122-state estimator and repeating the SCALE ablation at the identical noise realizations gives a correlation between the persistent parameters of corr($\hat{\gamma}$, $\hat{s}_p$) = −0.565338 and a condition number of 3.6013 for the unit-normalized persistent information block D M D, D = diag($M_{jj}^{-1/2}$). Normalization is necessary because the $\gamma$ and $s_p$ columns differ in magnitude by roughly three orders of magnitude; the condition number of the raw block reports the choice of units rather than the geometry. The two persistent modes are therefore separable at the baseline stellar field, not marginally so.

The consequence is a complete repair of the dominant failure mode. Under the SCALE ablation the 121-state estimator returns a bias of $+3.804477 \times 10^{-2}$ in $\gamma$, a covariance-consistency ratio $\eta_\gamma = 7126.6$, and zero coverage at both 1σ and 2σ. The 122-state estimator on the same realizations returns a bias of $+6.567987 \times 10^{-6}$, $\eta_\gamma = 1.0306$, and coverage of 0.652 and 0.952. The bias falls by a factor of $5.8 \times 10^3$, from 124 formal standard deviations to 0.018. Against the Monte Carlo standard errors at $N_{MC} = 1000$ — 2.2 per cent relative on $\eta_\gamma$, 0.015 on the 1σ coverage fraction, 0.007 on the 2σ fraction — the augmented estimator is statistically consistent with its own covariance. The plate-scale state is itself recovered essentially without bias: $-6.242736 \times 10^{-10}$ against an empirical dispersion of $2.870609 \times 10^{-8}$, with an empirical-to-formal ratio of 1.015. The $3.0 \times 10^{-4}$ injected calibration error is thus determined to about one part in $10^4$ of its prior width.

Figure 10 shows the effect directly. The 121-state normalized residual is centered — its bias is 1.7 per cent of its own sample dispersion — but spans four orders of magnitude in units of the formal standard deviation, so the failure is one of variance rather than of accuracy: the estimator is not systematically wrong, it is catastrophically overconfident. The 122-state residual is indistinguishable from a standard normal distribution.

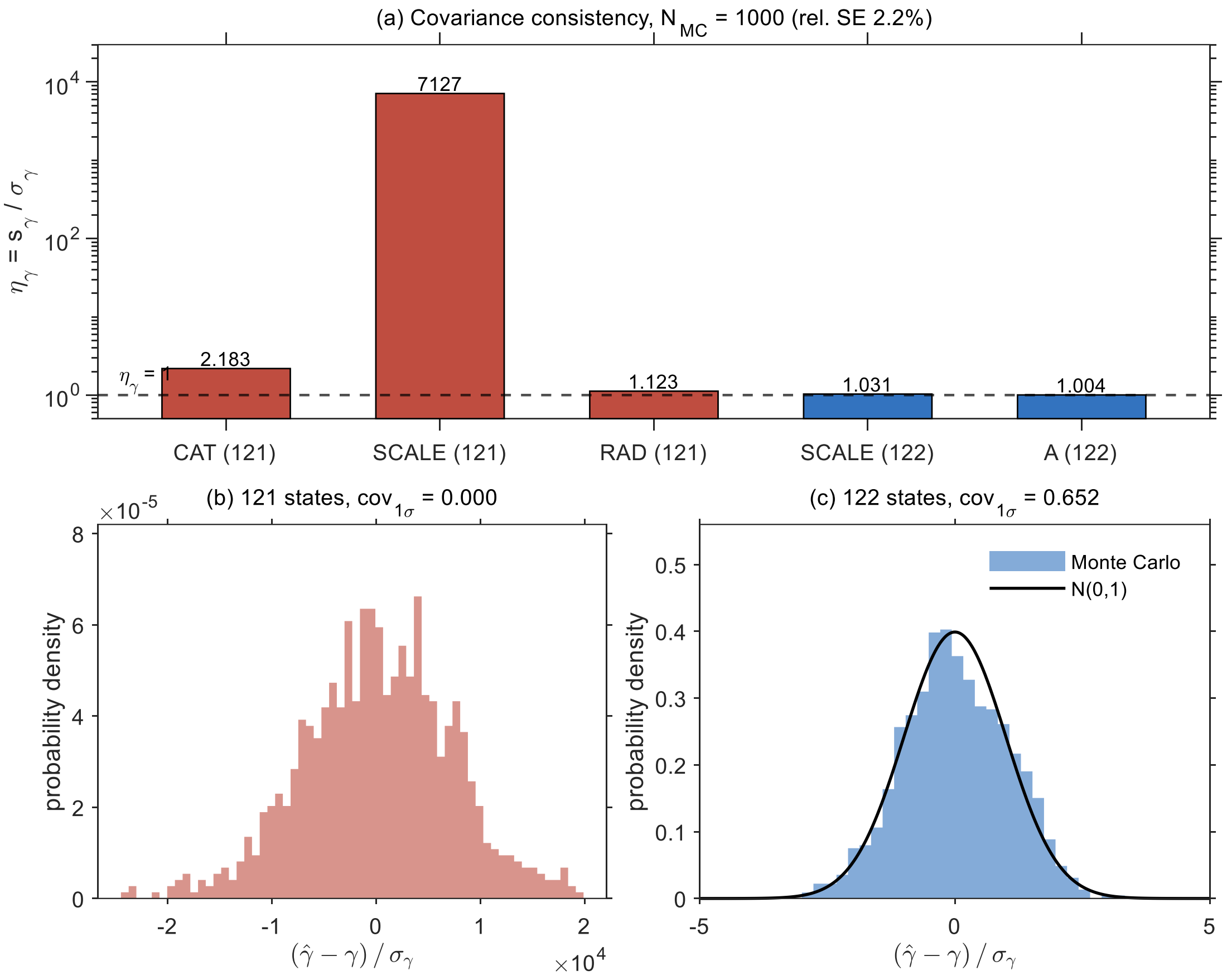

**Fig. 10. Effect of augmenting the state vector with plate scale, $N_{MC}$ = 1000. (a) Covariance-consistency ratio $\eta_\gamma = s_\gamma/\bar{\sigma}_\gamma$ for the three source-isolation ablations at 121 states and for the SCALE ablation and nominal Case A at 122 states; the dashed line marks $\eta_\gamma = 1$ and the vertical axis is logarithmic. Panel (a) extends Fig. 7 with the two augmented-estimator cases. (b) Normalized error $(\hat{\gamma} - \gamma)/\sigma_\gamma$ for the SCALE ablation with the 121-state estimator. Note the abscissa multiplier: the distribution is centered but spans $\pm 2 \times 10^4$ formal standard deviations. (c) The same quantity with the 122-state estimator of Eq. (129), plotted on the interval ±5 with the standard normal density overlaid for comparison. Panels (b) and (c) use identical noise realizations.**

Table 3 collects the comparison. The cost is a fixed inflation of the γ uncertainty, from $\sigma_\gamma = 3.075104 \times 10^{-4}$ to $3.728033 \times 10^{-4}$. This is not an empirical result but an identity: the marginal uncertainty of a parameter after a second parameter is estimated jointly is set entirely by their correlation, $\sigma_\gamma^{aug}/\sigma_\gamma = (1 - \mathrm{corr}^2(\hat{\gamma}, \hat{s}_p))^{-1/2} = 1.2123$, which reproduces the observed ratio 1.2123273 to better than one part in $10^6$. Carrying the calibration state costs 21.2 per cent in γ precision and nothing else. Repeating the augmented run on Case A, where the plate-scale error is at its nominal amplitude rather than the ablation value, gives $\eta_\gamma$ = 1.0043 with coverage 0.687 and 0.950, confirming that the extra state does not destabilize the nominal configuration; the 21.2 per cent premium is paid whether or not a large calibration error is present, because the formal covariance depends on the stellar geometry alone.

The augmented estimator therefore converts plate scale from an unmodeled systematic that invalidates the covariance into a jointly estimated calibration parameter at a bounded and predictable cost. Selected catalog corrections or low-order distortion states could be added by the same construction, and the correlation identity generalizes: the price of each additional persistent calibration state is set by its correlation with $\gamma$, so the design question for a flight instrument is not whether to estimate calibration jointly but which modes are sufficiently decorrelated from the deflection signature to be worth carrying.

*Table 3. Covariance consistency of the 121-state estimator of Eq. (23) and the 122-state augmented estimator of Eq. (129), evaluated on identical noise realizations. $\eta_\gamma = s_\gamma/\bar{\sigma}_\gamma$ is the ratio of the Monte Carlo sample dispersion to the mean formal standard deviation; $\eta_\gamma = 1$ indicates a covariance consistent with the realized errors. Nominal coverage is 0.6827 and 0.9545.*

| **Case** | **States** | **Bias $\gamma$** | **$s_\gamma$** | **$\bar{\sigma}_\gamma$** | **$\eta_\gamma$** | **cov 1σ** | **cov 2σ** |
|---|---|---|---|---|---|---|---|
| SCALE | 121 | $+3.804477 \times 10^{-2}$ | 2.191498 | $3.075104 \times 10^{-4}$ | 7126.6 | 0.000 | 0.000 |
| SCALE | 122 | $+6.567987 \times 10^{-6}$ | $3.842026 \times 10^{-4}$ | $3.728033 \times 10^{-4}$ | 1.0306 | 0.652 | 0.952 |
| A | 121 | $+6.274384 \times 10^{-6}$ | $3.039710 \times 10^{-4}$ | $3.075104 \times 10^{-4}$ | 0.9885 | 0.693 | 0.951 |
| A | 122 | $+1.089680 \times 10^{-5}$ | $3.744134 \times 10^{-4}$ | $3.728033 \times 10^{-4}$ | 1.0043 | 0.687 | 0.950 |

Plate-scale state (122-state runs): bias $-6.242736 \times 10^{-10}$, empirical σ $2.870609 \times 10^{-8}$, formal σ $2.828026 \times 10^{-8}$, 1σ coverage 0.660. Persistent block: corr($\hat{\gamma}$, ŝ_p) = −0.565338, cond(D M D) = 3.6013. $N_{MC}$ = 1000 per case; zero solver failures throughout.

### 5.8 Relationship to Existing Astrometric Estimation

That an unmodeled systematic degrades a formal covariance is a general property of misspecified least-squares estimation and is not itself new. What the ablation establishes is specific: the severity ordering across the three mismatch classes is not predictable from their amplitudes but follows from the radial structure of each error mode relative to the deflection signal, the dominant class exceeds covariance consistency by more than three orders of magnitude while every realization converges and no solution-internal diagnostic indicates failure, and the responsible quantity is an instrument calibration parameter rather than a term in the relativistic model. The consequence for mission design is that plate-scale knowledge belongs in the science error budget as an estimated state, since the calibration accuracy that would make it negligible is unattainable.

Gaia's global astrometric solution demonstrates the importance of solving jointly for source, attitude, calibration, and global parameters [8-10]. ECLIPSE-X is consistent with that broader principle but examines a narrower near-Sun multi-frame problem in which one relativistic parameter persists while navigation is frame local. The novelty is therefore not WLS, Fisher information, Schur-complement marginalization, PPN theory, or simultaneous astrometric calibration by themselves. It is the formulation and numerical characterization of this persistent-$\gamma$/frame-local-navigation architecture and the quantified identification of hidden plate-scale error as the dominant tested model-mismatch failure mode.

For space physics and relativistic astrometry, this result is useful because it identifies a concrete boundary between physical inference and instrument/navigation calibration. Solar gravitational deflection is not treated only as a correction applied after navigation, nor is spacecraft attitude assumed known a priori; the relativistic signal and frame-local navigation are separated within one estimation architecture. The ablation results then show which unmodeled astrometric mode most strongly compromises that separation in the tested configuration. Thus, the practical contribution is a validated framework for asking whether a weak persistent physical signal can be distinguished from spacecraft motion and calibration structure before a high-precision relativistic result is interpreted as physically meaningful.

**5.9 Plate-Scale Calibration Tolerance**

The plate-scale ablation of Sec. 5.4 establishes severe sensitivity at the tested standard deviation

$$\sigma_p = 3.0 \times 10^{-4} \tag{131}$$

but a calibration requirement cannot be inferred from a single perturbation amplitude, and the augmented estimator of Sec. 5.7 removes the failure at that amplitude without by itself establishing one.

The tolerance itself follows from a sweep in $\sigma_p$ at fixed noise realizations. Without a plate-scale state the unmodeled error aliases onto $\gamma$ with a fixed geometric gain, so the covariance-consistency ratio takes the one-parameter form $\eta_\gamma(\sigma_p) = [1 + (\sigma_p/\sigma_p^*)^2]^{1/2}$, where $\sigma_p^*$ is the amplitude at which the aliased contribution equals the formal uncertainty. The scale is obtained by least-squares regression of $\eta_\gamma^2 - 1$ on $\sigma_p^2$ through the origin. Repeating the SCALE ablation with the 121-state estimator across $\sigma_p$ from 0 to $1.0 \times 10^{-6}$, at $N_{MC} = 1000$ identical noise realizations per point and with no solver failures at any amplitude, gives $\sigma_p^* = 4.2037 \times 10^{-8}$ and reproduces the measured ratios to 1.13 per cent over three decades (Fig. 11a). The estimator crosses $\eta_\gamma = 1.2$ at $\sigma_p = 2.7884 \times 10^{-8}$. Its $1\sigma$ coverage leaves the ±1 Monte Carlo standard error band about 0.6827 at $\sigma_p = 1.0525 \times 10^{-8}$, so coverage is the more sensitive of the two diagnostics by a factor of 2.65, although at this ensemble size that crossing corresponds to $\eta_\gamma = 1.0309$ and is only marginally resolved (Fig. 11b). At the three largest amplitudes the bias in $\hat{\gamma}$ is 1.76, 1.79 and 1.80 per cent of the sample dispersion, so the failure is one of variance at every amplitude, as in Sec. 5.7. A mission that declines the extra state must therefore know its plate scale to three parts in $10^8$, a factor of $1.1 \times 10^4$ tighter than the amplitude treated as representative in Sec. 5.4. The augmented estimator is not an optional refinement but the only route to a trustworthy covariance at attainable calibration accuracy, and its 21.2 per cent precision premium should be read against that alternative.

One question remains open: whether that premium can be reduced by stellar-field design, since corr($\hat{\gamma}$, $\hat{s}_p$) depends on the realized distribution of impact parameters and is therefore a quantity the observing geometry can be optimized against.

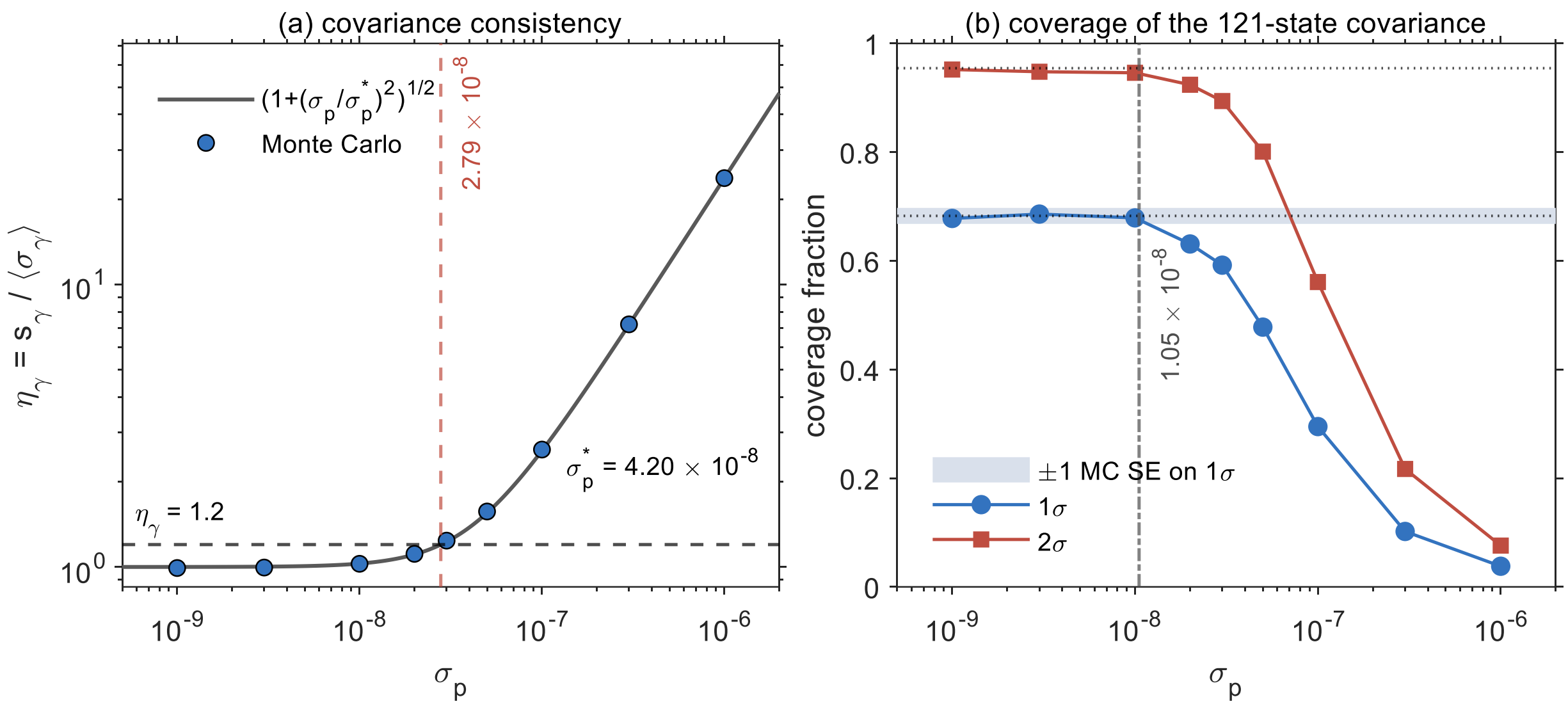

**Fig. 11. Plate-scale calibration tolerance of the 121-state estimator, $N_{MC}$ = 1000 per point at identical noise realizations. (a) Covariance-consistency ratio $\eta_\gamma$, the sample dispersion of $\hat{\gamma}$ divided by the mean formal uncertainty, against calibration uncertainty $\sigma_p$, with the one-parameter model $[1 + (\sigma_p/\sigma_p^*)^2]^{1/2}$ overlaid at the fitted $\sigma_p^* = 4.2037 \times 10^{-8}$; the dashed horizontal line marks the covariance-consistency criterion $\eta_\gamma = 1.2$ and the dashed vertical line its crossing at $\sigma_p = 2.7884 \times 10^{-8}$. (b) 1σ and 2σ coverage fractions against $\sigma_p$; dotted lines mark the nominal 0.6827 and 0.9545, the shaded band is ±1 Monte Carlo standard error on the 1σ fraction, and the dash-dotted vertical line marks where the 1σ fraction leaves that band, at $\sigma_p = 1.0525 \times 10^{-8}$. The $\sigma_p = 0$ point, at $\eta_\gamma = 0.9902$, cannot be shown on either logarithmic abscissa.**

### 5.10 Limitations and Future Work

The present study is a controlled numerical investigation rather than a flight-performance prediction. The stellar catalog is synthetic, and its radial sampling law is designed for estimator testing rather than astrophysical population fidelity. The values 1.22-8 apparent solar radii, 250 stars, 40 frames, 5-s cadence, navigation trajectories, centroid uncertainty, and perturbation amplitudes define the baseline experiment and are not claimed to be globally optimal.

This applies with particular force to the two persistent-parameter results. The 21.2 per cent premium and the tolerance $\sigma_p = 2.7884 \times 10^{-8}$ are values obtained at the baseline field (Secs. 5.7 and 5.9), and both are governed by corr($\hat{\gamma}$, $\hat{s}_p$), which depends on the realized distribution of impact parameters. A field spanning a narrower range in b yields a larger premium and a tighter tolerance; the relation, not the value, is the transferable result.

Only leading-order solar PPN deflection is represented. A flight-level implementation would require a more complete relativistic astrometric model incorporating observer motion, aberration, stellar proper motion and parallax, additional gravitating bodies, and higher-order terms as appropriate [6,7]. Optical distortion, point-spread-function evolution, detector effects, thermal deformation, chromatic response, solar stray light, and correlated centroid errors are also not modeled comprehensively.

Future work will systematically vary the number of observed stars, number of frames, observation cadence, stellar-field geometry, and measurement precision, to quantify sensitivity, identify performance boundaries, and translate the present observability results into mission-level instrument and navigation requirements.

## 6. Conclusions

ECLIPSE-X was developed to investigate simultaneous recovery of solar gravitational light deflection and spacecraft navigation from localized multi-frame stellar astrometry. The estimator uses one persistent PPN parameter $\gamma$ together with independent LOS and roll states for each observation frame.

For a deterministic 250-star field observed over 40 frames, the nominal estimator recovered

$$\hat{\gamma} = 1.0001198167 \tag{132}$$

with navigation-marginalized formal uncertainty

$$\sigma_\gamma = 3.075104 \times 10^{-4} \tag{133}$$

and normalized error

$$\frac{\hat{\gamma} - \gamma_0}{\sigma_\gamma} = 0.3896 \tag{134}$$

where the nominal estimate lies only $0.3896\sigma$ from the GR value, inside its predicted $1\sigma$ interval.

A 1000-realization matched-model ensemble independently confirmed statistical calibration:

$$s_{\gamma,A} = 3.039710 \times 10^{-4}, \qquad \eta_{\gamma,A} = 0.9885 \tag{135}$$

$$C_{1,A} = 0.693, \qquad C_{2,A} = 0.951 \tag{136}$$

Hidden astrometric model discrepancy fundamentally changed this behavior. The combined mismatch environments produced

$$s_{\gamma,B} = 1.549842, \qquad s_{\gamma,C} = 2.224386 \tag{137}$$

despite formal uncertainties remaining near $3.2 \times 10^{-4}$.

Source-isolation ablation established the relative impact of the three tested structured perturbations:

$$\eta_{\gamma,\mathrm{RAD}} = 1.1228, \qquad \eta_{\gamma,\mathrm{CAT}} = 2.1835, \qquad \eta_{\gamma,\mathrm{SCALE}} = 7126.6 \tag{138}$$

Plate-scale mismatch alone produced

$$s_{\gamma,\mathrm{SCALE}} = 2.191498 \tag{139}$$

with

$$C_{1,\mathrm{SCALE}} = 0, \qquad C_{2,\mathrm{SCALE}} = 0 \tag{140}$$

and substantial degradation of the recovered navigation states. All numerical realizations nevertheless converged successfully. The experiments therefore establish

$$\text{numerical convergence} \;\not\Rightarrow\; \text{statistical validity} \tag{141}$$

For the ECLIPSE-X architecture examined here, plate-scale knowledge is a critical observability requirement. Plate scale must either be calibrated sufficiently well independently or incorporated explicitly into the estimation state before high-precision relativistic inference can be considered statistically reliable.

Absent an estimated plate-scale state, the 121-state estimator remains covariance-consistent only for $\sigma_p$ below $2.7884 \times 10^{-8}$; carrying the state costs 21.2 per cent in the precision of $\hat{\gamma}$ and is the only route to a trustworthy covariance at attainable calibration accuracy.

The present work therefore establishes both the feasibility of persistent-$\gamma$/frame-local-navigation estimation under matched conditions and the principal tested calibration vulnerability that must be addressed in subsequent estimator development.

**Declaration of competing interest**

The author declares no known competing financial interests or personal relationships that could have appeared to influence the work reported in this paper.

**Data and code availability**

The ECLIPSE-X MATLAB implementation (v3.2) used to generate every numerical result in this paper, together with the configuration files defining Cases A, B, and C and the CAT, SCALE, and RAD ablations, is available from the author on reasonable request. The study uses no observational data; all stellar geometry is synthetic and reproducible from the seed rng(14018,'twister') reported in Eq. (7).

## Funding

This work received partial institutional support from the Department of Applied Aviation Sciences at Embry-Riddle Aeronautical University. No external grant funding was used for this study.

**Acknowledgments**

The author would like to thank the Applied Aviation Sciences Department at Embry-Riddle Aeronautical University for partially supporting this work.

**CRediT Statement**

Pedro J. Llanos: Conceptualization, Methodology, Software, Formal analysis, Investigation, Writing – original draft, Writing – review & editing, Visualization.

**Nomenclature**

ASR — Advances in Space Research

ECLIPSE-X — name of the numerical relativistic-navigation framework used in this study

GAME — Gamma Astrometric Measurement Experiment

GR — general relativity

LATOR — Laser Astrometric Test of Relativity

LOS — line of sight

MC — Monte Carlo

PPN — parametrized post-Newtonian

WLS — weighted least squares

$\gamma$ — PPN space-curvature parameter

$\sigma_\gamma$ — formal standard uncertainty of $\gamma$

$s_\gamma$ — empirical Monte Carlo standard deviation of $\gamma$ estimates

$\eta_\gamma$ — ratio of empirical to mean formal $\gamma$ uncertainty

CAT — catalog-coordinate mismatch ablation; independent realization-wide two-dimensional star-coordinate errors hidden from the estimator

SCALE — plate-scale mismatch ablation; one realization-wide fractional expansion or contraction of the truth stellar field hidden from the estimator

RAD — frame-coherent radial mismatch ablation; one scalar radial displacement per frame applied along each star's local Sun-to-star radial direction

ppm — parts per million; 300 ppm = 3.0 × 10^-4 fractional scale uncertainty

sigma_cent — one-standard-deviation stellar centroid-location uncertainty in detector pixels

sigma_theta — one-standard-deviation angular measurement uncertainty used by the estimator

sigma_p — standard deviation of the fractional plate-scale perturbation

sigma_cat — standard deviation of each catalog-coordinate perturbation component

sigma_r — standard deviation of the frame-coherent radial perturbation

$b_i$ — physical solar impact parameter of the light ray from star i

$q_i$ — angular separation of star i from the Sun expressed in apparent solar radii, $q_i = \frac{\rho_i}{\rho_\odot}$

q — MATLAB variable q — distinct geometric factor $\frac{2GM_\odot}{c^2 b_i}$ of Eq. (18); not the same quantity as $q_i$

$n_k$ — frame-local navigation state $[\delta\theta_{x,k}, \delta\theta_{y,k}, \delta\psi_k]^T$

g — stacked measurement sensitivity vector with respect to gamma

A — frame-local navigation design/Jacobian matrix

F — Fisher information matrix

## Declaration of generative AI and AI-assisted technologies in the manuscript preparation process

During the preparation of this work the author used OpenAI ChatGPT and Academia Co-scientist to assist with language editing and manuscript organization. The author reviewed and edited the content and takes full responsibility for the content of the published article.

**References**

[1] Dyson, F.W., Eddington, A.S., Davidson, C., 1920. A determination of the deflection of light by the Sun's gravitational field, from observations made at the total eclipse of May 29, 1919. Philos. Trans. R. Soc. Lond. A 220, 291-333.

[2] Will, C.M., 1993. Theory and Experiment in Gravitational Physics, 2nd ed. Cambridge University Press, Cambridge.

[3] Will, C.M., 2014. The confrontation between general relativity and experiment. Living Rev. Relativ. 17, 4.

[4] Misner, C.W., Thorne, K.S., Wheeler, J.A., 1973. Gravitation. W.H. Freeman, San Francisco.

[5] Bertotti, B., Iess, L., Tortora, P., 2003. A test of general relativity using radio links with the Cassini spacecraft. Nature 425, 374-376.

[6] Klioner, S.A., 2003. A practical relativistic model for microarcsecond astrometry in space. Astron. J. 125, 1580-1597.

[7] Soffel, M., Klioner, S.A., Petit, G., et al., 2003. The IAU 2000 resolutions for astrometry, celestial mechanics and metrology in the relativistic framework: explanatory supplement. Astron. J. 126, 2687-2706.

[8] Gaia Collaboration, Prusti, T., de Bruijne, J.H.J., Brown, A.G.A., et al., 2016. The Gaia mission. Astron. Astrophys. 595, A1.

[9] Lindegren, L., Lammers, U., Hobbs, D., O'Mullane, W., Bastian, U., Hernández, J., 2012. The astrometric core solution for the Gaia mission: overview of models, algorithms, and software implementation. Astron. Astrophys. 538, A78.

[10] Hobbs, D., Holl, B., Lindegren, L., Raison, F., Klioner, S.A., Butkevich, A., 2010. Determining PPN gamma with Gaia's astrometric core solution. IAU Symp. 261, 315-319.

[11] Turyshev, S.G., Shao, M., Nordtvedt, K.L. Jr., 2004. Experimental design for the LATOR mission. Int. J. Mod. Phys. D 13, 2035-2064.

[12] Turyshev, S.G., Shao, M., Nordtvedt, K.L. Jr., 2007. Mission design for the laser astrometric test of relativity. Adv. Space Res. 39, 297-304.

[13] Gai, M., Vecchiato, A., Lattanzi, M.G., Ligori, S., Loreggia, D., 2009. Gamma Astrometric Measurement Experiment (GAME) -implementation and performance. Adv. Space Res. 44, 588-596.

[14] Vecchiato, A., Gai, M., Lattanzi, M.G., Crosta, M., Sozzetti, A., 2009. Gamma Astrometric Measurement Experiment (GAME) -science case. Adv. Space Res. 44, 579-587.

[15] Kay, S.M., 1993. Fundamentals of Statistical Signal Processing, Volume I: Estimation Theory. Prentice Hall.

[16] Bar-Shalom, Y., Li, X.R., Kirubarajan, T., 2001. Estimation with Applications to Tracking and Navigation. Wiley.

[17] Crassidis, J.L., Junkins, J.L., 2012. Optimal Estimation of Dynamic Systems, 2nd ed. CRC Press.

[18] Zhang, F. (Ed.), 2005. The Schur Complement and Its Applications. Springer.